\documentclass[useAMS,usenatbib]{mn2e}

\usepackage[dvips]{graphicx}
\usepackage{amsmath,amssymb}
\usepackage{color}

\title[Eventful Evolution of GMCs in Dynamic Spirals]{
Eventful Evolution of Giant Molecular Clouds in Dynamically Evolving Spiral Arms
}
\author[J. Baba et al.]
{
Junichi \textsc{Baba}$^{1,2} $\thanks{E-mail: baba@cosmos.phys.sci.ehime-u.ac.jp; babajn2000@gmail.com}, 
Kana \textsc{Morokuma-Matsui}$^{3,4}$,
and 
Takayuki R. \textsc{Saitoh}$^1$
\\
$^1$ Earth-Life Science Institute, Tokyo Institute of Technology, 
2--12--1 Ookayama, Meguro-ku, Tokyo 152--8551, Japan.\\
$^2$ Research Center for Space and Cosmic Evolution, Ehime University, 
2--5 Bunkyo-cho, Matsuyama 790-8577, Japan\\
$^3$ Nobeyama Radio Observatory, National Astronomical Observatory of Japan, 
462--2 Nobeyama, Minamimaki-mura, Minamisaku-gun,\\ Nagano 384--1305, Japan.\\
$^4$ Chile Observatory, National Astronomical Observatory of Japan, 
2--21--1 Osawa, Mitaka, Tokyo 181--8588, Japan.
}

\begin{document}

\date{Accepted 2016 September 15. Received 2016 September 14; in original form 2016 March 7}


\maketitle

\begin{abstract}
The formation and evolution of giant molecular clouds (GMCs) in spiral galaxies have been investigated 
in the traditional framework of the combined quasi-stationary density wave and galactic shock model. 
In this study, we investigate the structure and evolution of GMCs 
in a dynamically evolving spiral arm using a three-dimensional $N$-body/hydrodynamic simulation 
of a barred spiral galaxy at parsec-scale resolution.
This simulation incorporated self-gravity, molecular hydrogen formation, radiative cooling, 
heating due to interstellar far-ultraviolet radiation, and stellar feedback by both ${\rm H_{II}}$ regions and Type-II supernovae. 
In contrast to a simple expectation based on the traditional spiral model, 
the GMCs exhibited no systematic evolutionary sequence across the spiral arm. 
Our simulation showed that the GMCs behaved as highly dynamic objects 
with eventful lives involving collisional build-up, collision-induced star formation, and destruction via stellar feedback.
The GMC lifetimes were predicted to be short, only a few tens of millions years.
We also found that, at least at the resolutions and with the feedback models used in this study, 
most of the GMCs without ${\rm H_{II}}$ regions were collapsing, but half of the GMCs with ${\rm H_{II}}$ regions 
were expanding owing to the ${\rm H_{II}}$-region feedback from stars within them.
Our results support the dynamic and feedback-regulated GMC evolution scenario.
Although the simulated GMCs were converging rather than virial equilibrium, 
they followed the observed scaling relationship well.
We also analysed the effects of galactic tides and external pressure on GMC evolution 
and suggested that GMCs cannot be regarded as isolated systems
since their evolution in disc galaxies is complicated because of these environmental effects.
\end{abstract}

\section{Introduction}
\label{sec:Introduction}

Understanding star formation mechanisms within galaxies 
is one of the most fundamental issues regarding the evolution of galaxies.
It is known that the majority of stars form in giant molecular clouds (GMCs), 
which have masses of about $10^{4}$--$10^7~\rm M_\odot$ and sizes of about 10--100 pc \citep[e.g.][]{Colombo+2014}.
Recent observational studies have shown that star formation activity \citep[e.g.][]{Muraoka+2009,Momose+2010,Hirota+2014,HuangKauffmann2015},
as well as the properties of the interstellar medium (ISM) and GMCs \citep[e.g.][]{Koda+2009,Koda+2012,Rebolledo+2012,
Schinnerer+2013,Hughes+2013,Colombo+2014,Muraoka+2016}, 
vary for different galactic structures such as spirals, inter-arms, and bars.
Accordingly, understanding the formation and evolution of GMCs, 
as well as the relationships between these processes and galactic structures, 
is a key factor in building a complete picture of star formation in galaxies.
In this paper, the first of this series, 
we focus on the formation, evolution, and dynamical state of GMCs in spiral arms.

According to the traditional view of GMC evolution in spiral arm environments, 
GMC evolution is influenced by `galactic shock' \citep[][]{Fujimoto1968,Roberts1969,Shu+1973} 
owing to so-called `quasi-stationary density waves' \citep[][]{LinShu1964,BertinLin1996}:
if gas flows {\it across} a spiral arm, it experiences a sudden compression due to the galactic shock and is followed 
by cloud formation in a phase transition \citep{Shu+1972,Kim+2008}, shock compression of incoming clouds \citep{Woodward1976}, 
and gravitational collapse \citep{Elmegreen1979,KimOstriker2002}, which triggers star formation. 
This scenario predicts that GMCs have evolutionary sequences across spiral arms in a wide radial range,
which has been preferred by recent observations of GMCs in nearby spiral galaxies 
such as M51 \citep[e.g.][]{Egusa+2011} and IC 342 \citep{Hirota+2011}.

Several galactic-scale hydrodynamic simulations have been performed to investigate 
the dynamical response of the ISM to a {\it rigidly rotating} spiral potential, which models quasi-stationary density waves. 
Such hydrodynamic simulations have highlighted the roles of spiral arms in the formation and evolution of GMCs;
the spiral potentials would enhance cloud--cloud collisions, which occur every 8--10 Myr, and
result in the formation of massive GMCs up to $10^6$--$10^7~\rm M_\odot$ \citep{Dobbs2008,Dobbs+2015}. 
\citet{Dobbs+2011a} showed that these GMCs are predominantly gravitationally unbound objects, 
because cloud-cloud collisions and stellar feedback generate internal random motions of the gas.
In addition to GMC formation, hydrodynamic simulations in spiral potentials showed that 
when a GMC passes through a spiral arm, strong shear within the spiral arm contributes to 
the destruction of the GMC \citep{WadaKoda2004,DobbsBonnell2006}.
In particular, \citet{DobbsPringle2013} suggested that smaller $(\sim 10^4~\rm M_\odot$) clouds are strongly 
affected by feedback, whereas larger ($\sim 10^6~\rm M_\odot$) clouds are more affected by shear.

In contrast to the traditional quasi-stationary density wave theory, 
wherein stellar spiral arms slowly change during several galactic rotation periods \citep[i.e. $\gtrsim 1$ Gyr;][]{BertinLin1996}, 
an alternative spiral theory has been developed \citep[see][for a recent review]{DobbsBaba2014}.
This so-called `dynamic' spiral theory predicts that stellar spiral arms are dynamic structures, 
with amplitudes that change on the galactic rotation time scale, or even more rapidly 
\citep[i.e. a few hundred million years;][]{SellwoodCarlberg1984,Fujii+2011,Sellwood2011,
Grand+2012a,Baba+2013,D'Onghia+2013,Pettitt+2015}. 
Dynamic spirals are also observed in $N$-body barred spiral galaxies 
\citep{Baba+2009,Grand+2012b,Roca-Fabrega+2013,Baba2015c}.
In contrast to quasi-stationary density waves, such dynamic spirals {\it co-rotate} with material
at any given radius\footnote{
Co-rotational behabiour is not true for tidally induced spirals such as M51.
Recent numerical simulations of tidally interacting systems suggested that 
the pattern speeds of tidally induced spirals clearly differ from the galactic angular speed 
but that they decrease with increasing radius \citep{Dobbs+2010,Pettitt+2016}.
} \citep{Wada+2011,Grand+2012a,Baba+2013,Kawata+2014}.
This co-rotational behaviour may occur because the swing amplification is 
most efficient near the co-rotation radius \citep[e.g.][]{Toomre1981} and 
non-linear interactions between the swing-amplified wakelets develop 
a global spiral arm \citep{KumamotoNoguchi2016}. 
Furthermore, \citet{Wada+2011} and \citet{DobbsBonnell2008} performed hydrodynamic simulations 
of dynamic multiple spirals and found that the gas effectively falls into the spiral potential minimum 
from {\it both sides} of the spiral arm (the so-called `large-scale colliding flows'), 
rather than passing through the spiral arms, as predicted by galactic shock theory.
\citet{Baba+2016a} observed such large-scale colliding flow 
in a simulated barred galaxy with a dynamic grand-design spiral.

This recent progress related to the dynamics of spiral arms suggests that a rigidly rotating spiral could not be 
a reasonable dynamical model for investigating the formation and evolution of GMCs.
It is therefore necessary to investigate the differences between 
these processes in quasi-stationary density waves and dynamic spirals.
\citet{Renaud+2013} and \citet{Hopkins+2012} investigated GMCs by using $N$-body/hydrodynamic simulations
with sub-parsec-scale or parsec-scale resolutions, although they did not focus on the relationship between GMCs and spiral arms.
Thus, the formation and evolution of GMCs, as well as the relationship between GMCs and spiral arms, 
have not yet been well explored in a dynamic spiral context.

In this study, we performed a three-dimensional $N$-body/hydrodynamic simulation of a barred spiral galaxy 
at a parsec-scale resolution in order to investigate the structures, formation and evolution of GMCs in a galactic context.
Our galaxy model has {\it dynamic} barred grand-design spirals, which develop spontaneously 
from an axisymmetric bulge-disk-halo system \citep{Baba2015c}.
To study the effects of the dynamical behaviours of spiral arms,
we also performed a hydrodynamic simulation of a {\it fixed} spiral potential with a single pattern speed, 
which mimics the traditional spiral model.

This paper is structured as follows.
In Section \ref{sec:Method}, we describe our simulation methods regarding self-gravity, radiative cooling, 
heating due to interstellar radiation, molecular hydrogen formation, star formation from cold dense gas, 
and stellar feedback from $\rm H_{II}$ regions/supernova (SN) explosions.
We describe the differences between the molecular gas structures and GMCs around the two types of spiral arms in Section \ref{sec:Comparison}.
We then present the results regarding the formation, evolution, and dynamical state of GMCs in the dynamic spirals in Section \ref{sec:DynamicSpiralResults}.
In particular, we focus on one simulated GMC in the spiral region and describe its evolution in detail
and then present the statistical dynamical properties of GMCs, as well as the effects of galactic tides and external pressures.
The formation and evolution of GMCs in the fixed spiral model are presented in Section \ref{sec:FixedSpiralResults}.
Finally, we summarize the results in Section \ref{sec:Summary}.

\section{Numerical Method and Galaxy Models}
\label{sec:Method}

We used our original $N$-body/smoothed particle hydrodynamics (SPH) code {\tt ASURA-2}
\citep{SaitohMakino2009,SaitohMakino2010} to perform three-dimensional hydrodynamic simulations.
{\tt ASURA-2} implements a time-step limiter that allows us to solve rapid expansions 
by imposing sufficiently small time-step differences between neighbouring particles \citep{SaitohMakino2009}.
The FAST method, which accelerates the time integration of a self-gravitating fluid through the assignment of different time steps 
for gravity and the hydrodynamic interactions of each particle, was also implemented \citep{SaitohMakino2010}.
The self-gravities of the stellar and SPH particles were computed using the Tree with the GRAPE\footnote{
In this study, we used a software emulator of GRAPE, Phantom-GRAPE \citep{Tanikawa+2013}, 
which is available at https://bitbucket.org/kohji/phantom-grape.
} method \citep{Makino1991}.

\subsection{Gas Cooling and Heating}
\label{sec:CoolingHeating}

\begin{figure}
\begin{center}
\includegraphics[width=0.4\textwidth]{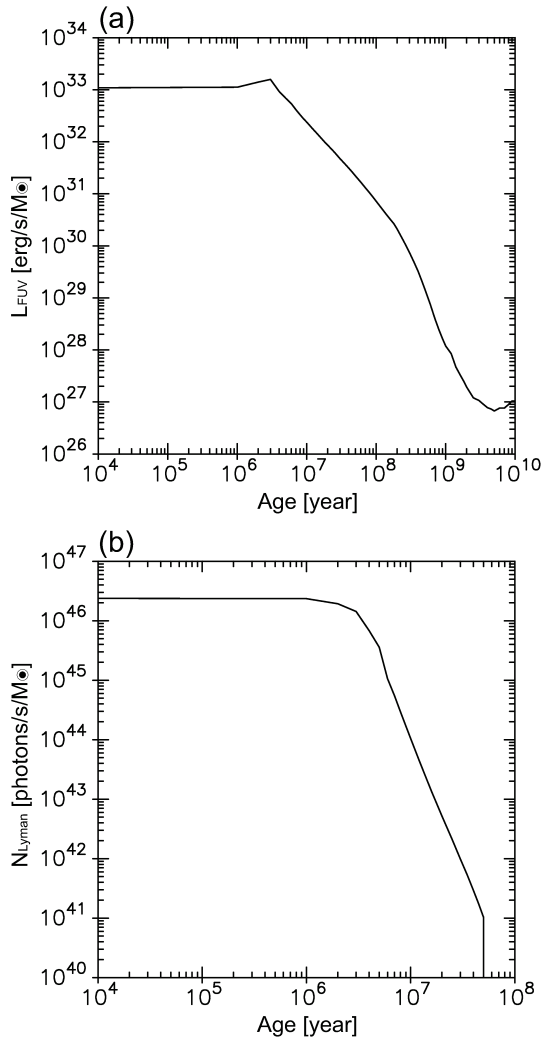}
\caption{
	(a) Specific FUV luminosity and (b) specific numbers of emitted ionization photons 
	as functions of the age of a stellar particle in SSP approximation.
	See Section \ref{sec:SF} for the parameters of SSP approximation.
}	
\label{fig:FUVluminositySSP}
\end{center}
\end{figure}

The radiative cooling of the gas was determined by assuming an optically thin cooling function, 
$\Lambda(T, f_{\rm H2} ,G_{\rm 0})$, based on a radiative transfer model of photo-dissociation regions 
for a wide temperature range of $20~{\rm K} < T < 10^8~{\rm K}$ \citep[see Appendix of][]{Wada+2009}.
In this cooling function, the molecular hydrogen fraction, $f_{\rm H2}$, follows the fitting formula given by \citet{GnedinKravtsov2011},
and $G_0$ is the far-ultraviolet (FUV) intensity normalized to the solar neighbourhood value.
We also included the photoelectric heating of the gas due to interstellar FUV radiation \citep[][]{Wolfire+1995}.
The heating rate was calculated using:
\begin{eqnarray}
 \Gamma = 10^{-24} \epsilon_{\rm PE} G_0~\rm ergs~s^{-1},
\end{eqnarray}
where $\epsilon_{\rm PE} $ is the fraction of FUV radiation absorbed by grains, which is converted to gas heating (i.e. heating efficiency).
Note that $\epsilon_{\rm PE}$ depends on the value of $G_0T^{1/2}/n_{\rm e}$ 
\citep[where $n_{\rm e}$ is the number density of electrons; see Fig. 1 of][]{Wolfire+1995} 
and decreases by a factor of 10 for $G_0T^{1/2}/n_{\rm e} \gtrsim 10^3$.
Although the high-$G_0T^{1/2}/n_{\rm e}$ situation corresponds to environments with high $G_0$, high $T$, or low density, 
SN heating or ${\rm H_{II}}$-region heating is predominant in these environments. 
Thus, for simplicity, we used $\epsilon_{\rm PE} = 0$ for gas with $T>10^4$ K
and $\epsilon_{\rm PE} = 0.05$ for $T \lesssim 10^4$ K which is the typical range for a neutral gas.

In the present study, we adopted different methods of evaluating $G_0$ depending on whether or not an SPH particle was heated 
by stellar feedback (SNe or $\rm H_{II}$-regions). For the SPH particles heated by stellar feedback, we used $G_0 = 10^3$
in order to force the SPH particles to have $f_{\rm H2} \ll 1$. 
For the non-heated SPH particles, we calculated $G_0$ using the following equation:
\begin{eqnarray}
G_{\rm 0} = G_{\rm 0,thin} e^{-\sigma_{1000} N_{\rm H}},
\end{eqnarray}
where $N_{\rm H}$ is the total column density of hydrogen, 
$\sigma_{1000} = 2 \times 10^{-21}~\rm cm^2$ is the effective cross-section for dust extinction 
at $\lambda = 1000~\rm \AA$ \citep{DraineBertoldi1996,GloverMacLow2007a},
and $G_{\rm 0,thin}$ is a normalized FUV intensity in optically thin limit.
In this study, $N_{\rm H}$ was estimated by using a Sobolev-like approximation \citep{Gnedin+2009}:
\begin{eqnarray}
 N_{\rm H} \approx n_{\rm H} \frac{\rho}{|\nabla \rho|},
\end{eqnarray}
where $n_{\rm H}$ and $\rho$ are the number density of hydrogen and the mass density of an SPH particle, respectively,
and the evaluation of $\nabla \rho$ is based on the standard SPH formulation.
In contrast, $G_{\rm 0,thin}$ is given by a summation over all stellar particles \citep{GerritsenIcke1997,Pelupessy+2006}:
\begin{eqnarray}
 G_{\rm 0,thin} = \frac{1}{1.6 \times 10^{-3}~{\rm erg~cm^{-2}~s^{-1}}} \sum_i \frac{m_{\star,i}L_{{\rm FUV},i}}{4\pi r_i^2},
\end{eqnarray}
where the numerical value of $1.6 \times 10^{-3}~{\rm erg~cm^{-2}~s^{-1}}$ 
is the observed FUV intensity in the solar neighbourhood \citep{Habing1968},
and $m_{\star,i}$, $L_{{\rm FUV},i}$, and $r_i$ are the mass, time-dependent specific FUV luminosity, 
and distance of the $i$-th stellar particle, respectively. The tree method was used to compute this summation. 
We determined $L_{{\rm FUV},i}$ of the $i$-th stellar particle by following its age 
by using the stellar population synthesis modelling software PEGASE \citep{FiocRocca-Volmerange1997}.  
Fig. \ref{fig:FUVluminositySSP}a shows the time evolution of $L_{{\rm FUV},i}$.

\subsection{Star Formation and Stellar Feedback}
\label{sec:SF}

We implemented a sub-grid model for star formation into our simulation as follows.
Each stellar particle was considered to be a simple stellar population (SSP) having its own age, 
metallicity and initial mass function (IMF). Throughout this study, we assumed that 
the IMF was the Salpeter type \citep{Salpeter1955}, 
which has lower and upper mass limits of 0.1 $\rm M_\odot$ and 100 $\rm M_\odot$, respectively.
If an SPH particle satisfied the criteria (1) $n > 10^3~{\rm cm^{-3}}$; 
(2) $T < 100~{\rm K}$; and (3) $\nabla\cdot\mathbf{v}<0$, 
it created star particles in a probabilistic manner following the Schmidt law
with a local dimensionless star formation efficiency, $C_{\ast}=0.033$ \citep{Saitoh+2008}.
It should be noted that the global star formation properties, such as the star formation rate
and Schmidt--Kennicutt relation \citep[][and references therein]{KennicuttEvans2012}, 
are insensitive to the adopted value of $C_{\ast}$
and are instead primarily controlled by the global evolution of the ISM from reservoir to dense gas, 
where stars are formed \citep[][]{Saitoh+2008,Hopkins+2011}.  
We also note that \citet{Dobbs+2011b} adopted a density criterion of $10^3~\rm cm^{-3}$ in 
hydrodynamic simulations at a mass resolution level similar to that considered in the present investigation.

Type-II SN feedback was included in the thermal energy, and the times at which 
the SN events occurred were determined probabilistically, as in \citet{Okamoto+2008}.
In our probabilistic treatment, all of the SN events in a stellar particle were assumed 
to occur simultaneously, as a single event. 
Assuming that stars with masses larger than $8~\rm M_\odot$ exploded as Type-II SN,
the probability of a stellar particle undergoing an SN explosion during a time-step $\Delta t$ is given by
\begin{eqnarray}
p_{\rm SN} = \frac{\int_\tau^{\tau+\Delta t} r_{\rm SN}(t') dt'}{\int_\tau^{\tau_8} r_{\rm SN}(t') dt'},
\end{eqnarray}
where $\tau$ is the age of a stellar particle, $r_{\rm SN}$ is the Type-II SN rate for the stellar particle,
and $\tau_8$ is the lifetime of an $8~\rm M_\odot$ star (i.e. $\simeq 30$ Myr). 
A random number $p$ was then generated to determine whether the stellar particle underwent a Type-II SN event during $\Delta t$.
If $p < p_{\rm SN}$, we smoothly distributed the mass and feedback energy that were expelled by the total Type-II SN 
between $\tau = 0$ and $\tau_8$ over the surrounding neighbour SPH particles. 
The energy of each Type II-SN explosion was assumed to be $10^{51}$ erg.

We incorporated the ${\rm H_{II}}$-region feedback as pre-SN feedback 
using a Stromgren volume approach, in which the gas around young stars that extend out to 
a radius sufficiently large for ionization balance is simply set to 
a temperature of $10^4$ K \citep[see also][]{Hopkins+2012,Renaud+2013}. In our Stromgren volume approach, 
we calculated the Stromgren radius, $R_s$, to satisfy the following equation, 
assuming that an ionization balance exists in a homogeneous medium:
\begin{eqnarray}
R_{\rm s} =\left(\frac{3 S_{\rm Lyn}}{4\pi \overline{n_{\rm H}}^2 \alpha_{\rm B}}\right)^{1/3},
\end{eqnarray}
where $S_{\rm Lyn}$, $\overline{n_{\rm H}}$, and $\alpha_{\rm B}$ are the number of emitted ionization photons per unit time, 
average number density of hydrogen within $R_{\rm s}$, and case-B recombination coefficient 
\citep[$2.6 \times 10^{-13}~\rm cm^3~s^{-1}$ for $T=10^4$ K;][]{OsterbrockFerland2006}, respectively.
We computed $S_{\rm Lyn}(\tau) = m_{\star} N_{\rm Lyman}(\tau)$ as a function of the age, 
$\tau$, of a stellar particle with a mass of $m_\star$, where $N_{\rm Lyman}$ is 
the specific number of ionisation photons, which was calculated using PEGASE (Fig. \ref{fig:FUVluminositySSP}b).
If a star cluster with a mass several times larger than $10^3~\rm M_\odot$ formed 
in a typical GMC (with $\overline{n_{\rm H}} \sim 10^2~\rm cm^{-3}$), 
we estimated $R_{\rm s}$ as being equal to a few parsecs, 
which is comparable to a typical smoothing size in such a dense environment.

\subsection{Galaxy Models}

\subsubsection{Dynamic spiral model}
\label{sec:DynamicModel}

The initial axisymmetric model was composed of  live stellar/gaseous discs, 
a live classical bulge, and a fixed dark matter halo.
Hereafter, we refer to this model as the `DYN' model.
We briefly summarise the galaxy model and refer the readers to \citet{Baba2015c} for further details.
The stellar disc follows an exponential profile: 
\begin{eqnarray}
 \rho_{\rm disc}(R,z) = \frac{M_{\rm d}}{4\pi R_{\rm d}^2 z_{\rm d}} 
 \exp\left(-\frac{R}{R_{\rm d}}\right){\rm sech}^2\left(\frac{z}{z_{\rm d}}\right), 
\end{eqnarray}
where $M_{\rm d}$, $R_{\rm d}$, and $z_{\rm d}$ are the total mass, scale-length, and scale-height of the stellar disc, respectively.
The gas disc also follows an exponential profile with a total mass of $M_{\rm d,g}$, scale-length of $R_{\rm d,g}$, 
and scale-height of $z_{\rm d,g}$. 
The classical bulge follows the Hernquist profile \citep{Hernquist1990}:
\begin{eqnarray}
 \rho_{\rm bulge}(r) = \frac{M_{\rm b}}{2\pi}\frac{a_{\rm b}}{r(r+a_{\rm b})^3},
\end{eqnarray}
where $M_{\rm b}$ and $a_{\rm b}$ are the total mass and scale-length of the bulge, respectively.
For the fixed dark matter halo, we adopted the Navarro-Frenk-White profile \citep{Navarro+1997}:
\begin{eqnarray}
 \rho_{\rm halo}(r) =  \frac{M_{\rm h}}{4\pi f_{\rm c}(C_{\rm h})}\frac{1}{r(r+a_{\rm h})^2},
\end{eqnarray}
where $M_{\rm h}$, $a_{\rm h}$, and $C_{\rm h}$ are the total mass, virial radius, and concentration parameter 
of the dark matter halo, respectively, and $f_{\rm c}(C_{\rm h}) = \ln(1+C_{\rm h}) - C_{\rm h}/(1+C_{\rm h})$.
The values of the model parameters are listed in Table 1.
This galaxy model had the same circular velocity curves for each component as those shown in Fig. 1 of \citet{Baba+2015a}.
We produced the initial conditions via Hernquist's method \citep{Hernquist1993} 
using an azimuth shuffling procedure to prevent emergence of a ring structure \citep{McMillanDehnen2007,Fujii+2011}.

The initial numbers of star and SPH particles in the simulation were $6 \times 10^6$ and $4 \times 10^6$, respectively,
and the particle masses were approximately 9000 M$_\odot$ and 3000 M$_\odot$, respectively. 
We investigated the dynamical evolution of the galaxy by solving the basic equations of the galactic model,
with a gravitational softening length of $10$ pc until $t=2.4$ Gyr, and 
we then split each particle into eight or four daughter particles (according to the mass of the parent particle) 
inheriting the particle type of the parent particle.
The particle splitting procedure was as follows:
the positions of the daughter particles were determined randomly within 
a radius $r_{\rm split} =  1.5$ pc around the parent particle 
while conserving the barycenter of the parent particle, and the tiny velocity fluctuation were added.
The velocity fluctuations were assigned by a Gaussian distribution with a dispersion of $0.5 \sqrt{G m/r_{\rm split}}$ 
(here $G$ and $m$ are the gravitational constant and the mass of the parent particle, respectively).
Note that the results do not depend on the details of the particle splitting procedure, 
because we analysed the GMCs in the resimulation 200 Myr after the splitting. 
The resultant numbers of stellar and SPH particles were approximately 50 million and 18 million, respectively, 
and the masses of the refined stellar and SPH particles were approximately $650~\rm M_\odot$.
We continued the simulation with the refined particle distribution,
by using a gravitational softening length of $3$ pc until $t=2.65$ Gyr.

Note that \citet{Bonnell+2013} and \citet{Dobbs2015} performed resimulations of 
galactic-scale hydrodynamic simulations of fixed spiral models with their particle-splitting methods 
and investigated the structure and evolution of GMCs at much higher resolutions
\citep{MacLachlan+2015,SmilgysBonnell2016,Duarte-CabralDobbs2016}.
\citet{Bonnell+2013} did not include stellar feedback or investigate GMC properties and evolution, 
which are the focuses of the present paper. 
\citet{Dobbs2015} included self-gravity and stellar feedback at a higher resolution than 
the models presented here, but covering a smaller region of the galaxy.
To understand cloud evolution within galaxies, these studies are complementary to each other.

\begin{table}
\caption{Model parameters of the dynamic spiral model.}
\begin{center}
\begin{tabular}{cc|c}
 Component		& Notation		&	Value \\
\hline
\hline
Stellar Disc 		& $M_{\rm d}$		& $4.53 \times 10^{10}$ $M_{\rm \odot}$\\
				& $R_{\rm d}$		& $2.8$ kpc\\
				& $z_{\rm d}$		& $410$ pc\\	
\hline
Gas Disc 			& $M_{\rm d,g}$	& $1.2 \times 10^{10}$ $M_{\rm \odot}$\\
				& $R_{\rm d,g}$	& $11.2$ kpc\\
				& $z_{\rm d,g}$		& $100$ pc\\
\hline
Classical Bulge 	& $M_{\rm b}$		& $3.6 \times 10^{10}$ $M_{\rm \odot}$\\
				& $a_{\rm b}$		& $0.788$ kpc\\
\hline
Dark Matter Halo 	& $M_{\rm h}$		& $1.26 \times 10^{12}$ $M_{\rm \odot}$\\
				& $R_{\rm h}$		& $280$ kpc\\
				& $C_{\rm h}$		& $11.2$ \\
\hline
\end{tabular}
\end{center}
\label{tbl:ModelParameters}
\end{table}%

\subsubsection{Fixed spiral model}
\label{sec:SteadyModel}

In order to study the effect of the dynamical behaviour of spiral arms,
we run a test model, in which the stellar disc is replaced with a time-independent spiral potential. 
Hereafter, we refer to this model as the `SDW' model \citep[see][for details]{Baba+2016a}.
In the SDW model, we imposed the rigidly rotating spiral potential, $\Phi_{\rm sp}(R,\phi,z;t)$,
into the static axisymmetric potential, $\Phi_{\rm 0}(R,z)$, which consisted of a stellar disk, a spherical bulge and halo. 
The pitch angle and co-rotation (CR) radius of the spiral potential are $25^\circ$ and 15 kpc, respectively.
The CR radius corresponds to a pattern speed of $\Omega_{\rm p} \simeq 16~\rm km~s^{-1}~kpc^{-1}$ in the SDW model.
The self-gravity of the gas, radiative cooling/heating, star formation 
and stellar feedback (Type-II SN and ${\rm H_{II}}$-region feedback) were also considered.

The initial number of SPH particles and the gravitational softening length were the same as those in the DYN model.
Similarly to in DYN model, we first investigated the evolution of the gas disc with a gravitational softening length of 10 pc until $t = 100$ Myr 
and then applied the particle-splitting method to this result.
The resultant masses of the particles and the gravitational softening length were the same as those in the DYN model. 
The simulation of the refined particle distribution were performed until $t = 350$ Myr.

\section{Distributions of Spurs and GMCs Around Spiral Arms}
\label{sec:Comparison}

In this section, we describe the differences between the properties of the spurs and GMCs in the DYN and SDW models.
For the sake of this comparison, effects of the bar should be considered in the DYN model.
However, \citet{Baba2015c} found that the effects of bars are negligible in the outer regions ($R>$ 1.5--2 bar radii).
We thus focus on the spiral arms regions with $R > 6$ kpc of the DYN model.

\begin{figure*}
\begin{center}
\includegraphics[width=0.85\textwidth]{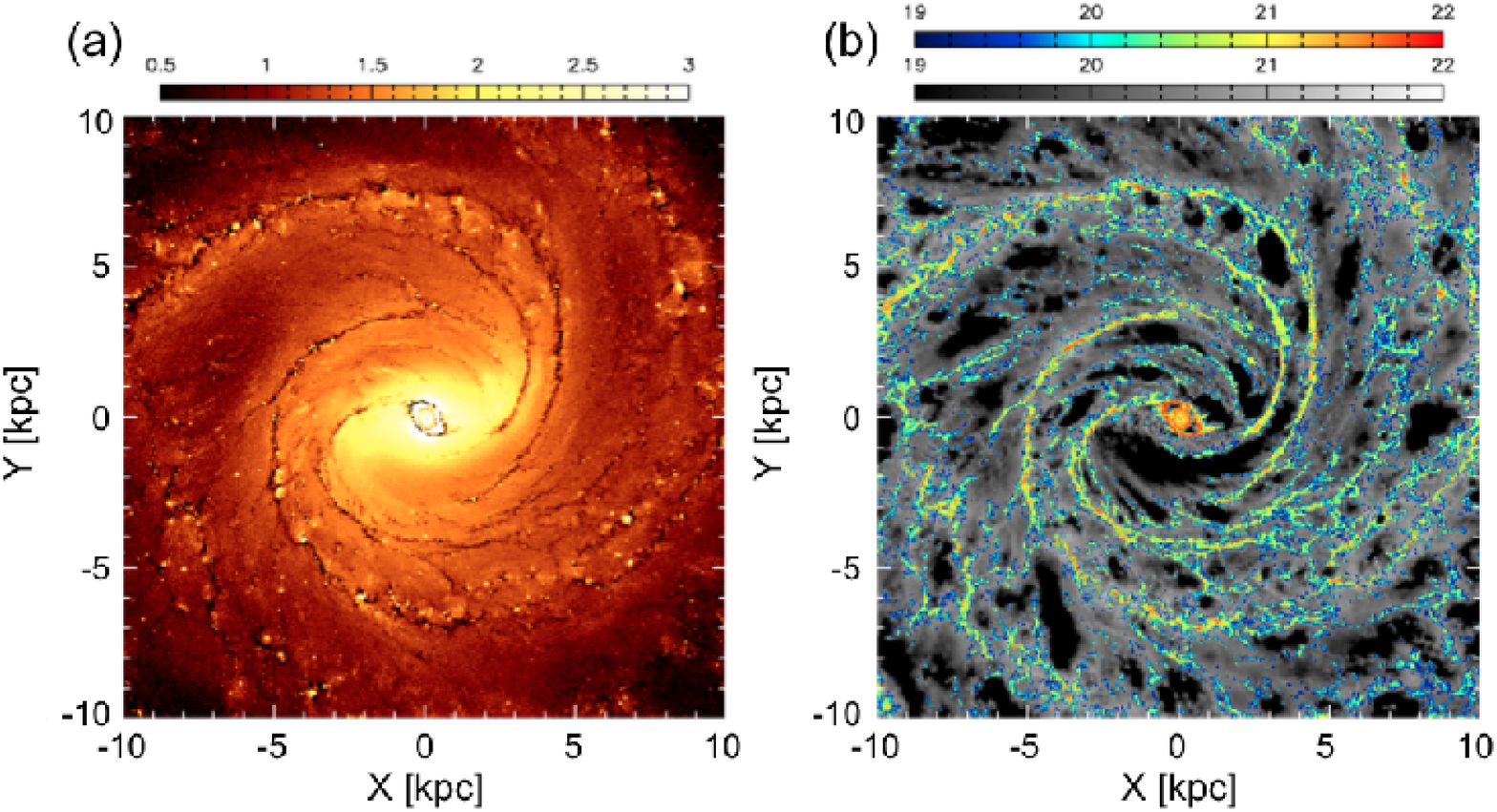}
\includegraphics[width=0.85\textwidth]{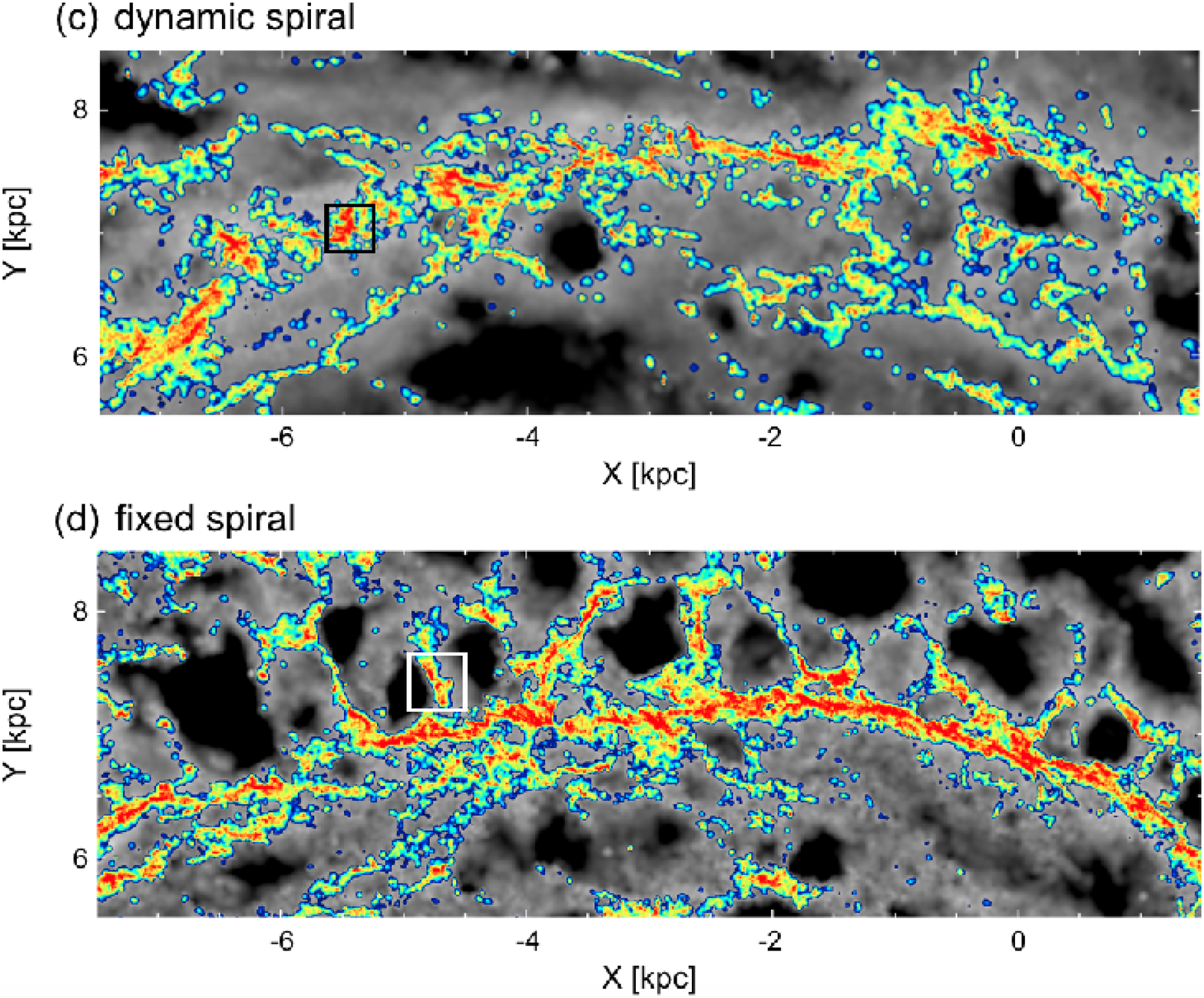}
\caption{
	(a) Face-on view of $B$-band map of DYN model
	with dust extinction in the unit of $\rm L_\odot~pc^{-2}$ (logarithmic scale) at $t = 2.625$ Gyr.
	In this model, the stellar bar developed as a result of bar instability at $t \simeq 1.5$ Gyr.
	The dust extinction for the $B$-band map was estimated through multiplication by a factor of $e^{-\tau_B}$, 
	where the optical depth $\tau_B$ was calculated from the absorption cross-section $\sigma_B = 6 \times 10^{-22}~\rm cm^2$ 
	and the total hydrogen column density $N_{\rm H}$. 
	(b) Same as panel (a), but for molecular (rainbow colours) and atomic (grey colours) gas 
	column densities in units of $\rm H~cm^{-2}$ (logarithmic scale).
	(c) Same as panel (b), but for a spiral arm region. 
	The formation and evolution of the GMC enclosed by the solid square are described in Section \ref{sec:GMCdetails}.
	(d) Same as panel (c), but for SDW model at $t = 340$ Myr. 
	Spurs (i.e. the shorter features which are perpendicular to the main spiral arms)
	appear in leading side of arm. Solid square encloses example of spurs.
}	
\label{fig:Snapshot}
\end{center}
\end{figure*}

\subsection{Spurs}
\label{sec:Spurs}

Figs. \ref{fig:Snapshot}a and \ref{fig:Snapshot}b show face-on views of 
the DYN model at $t = 2.6$ Gyr (i.e. 0.2 Gyr after the refinement). 
A stellar bar with a semi-major length of approximately 3 kpc 
and an associated stellar grand-design spiral arm are clearly observed.
The dark lanes are also evident along with the spiral arms and bar. 
As shown in Fig. \ref{fig:Snapshot}b, the atomic gas (${\rm H_I}$) is distributed throughout the galactic disc, 
whereas the molecular gas (${\rm H_2}$) is more localized than the stars and ${\rm H_I}$ gas. 
The molecular gas exhibits a complicated network of filaments and clumps and is associated with the stellar spiral arms.
The complicated structures of the gas are clearly seen in the zoomed-in map of a spiral region
shown in Fig. \ref{fig:Snapshot}c.

The complicated structures of the gas have been reported following studies in which hydrodynamic simulations 
of fixed spiral potentials were conducted \citep[e.g.][]{Wada2008,DobbsPringle2013,Dobbs2015},
although there is a clear difference between the DYN and SDW models.
Fig. \ref{fig:Snapshot}d shows a magnified view of the distributions of the gas in the spiral area of the SDW model.
In contrast to the DYN model, the SDW model shows clear spur features 
in the downstream (i.e. leading) side of the arm.
Such `single-side spurs' were reported by previous hydrodynamic simulations of fixed spiral potentials 
\citep[e.g.][]{WadaKoda2004,DobbsBonnell2006,KimOstriker2006,Wada2008,Dobbs2008,Pettitt+2016}.
This difference suggests that existence of `single-side spurs' in actual spiral galaxies 
can be used as an observational diagnostics of spiral structure theories.

\subsection{GMC Scaling Relations}
\label{sec:ScalingRelations}

To investigate the properties of GMCs, we defined the GMCs in our simulations as 
conjunct structures with molecular column densities greater than a threshold of
$3 \times 10^{21}~\rm H~cm^{-2}$ ($\simeq 30~\rm M_\odot pc^{-2}$)
in a face-on column density map of the molecular gas. 
We identified GMCs as follows:
first, the SPH particles were mapped onto a two-dimensional grid with a cell size of 5 pc to produce 
the face-on column density map of the molecular gas. Then, we identified conjunct structures  
using the Friend-of-Friend (FoF) method \citep[e.g.][]{Efstathiou+1985}, with a linking length of 10 pc.

\begin{figure}
\begin{center}
\includegraphics[width=0.48\textwidth]{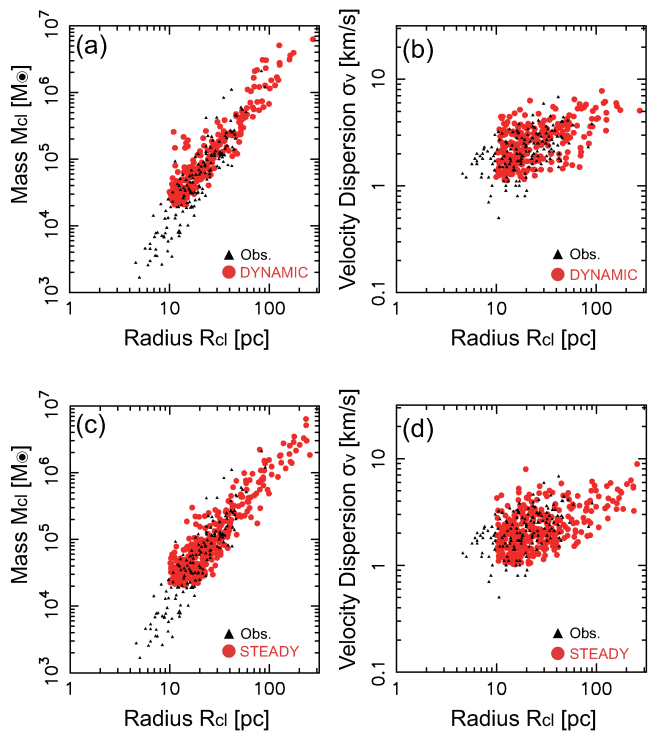}
\caption{
	Top panels: Scaling relations between mass, size, and velocity dispersions of observed GMCs 
	\citep[black triangles;][]{Heyer+2009} and simulated (red filled circles) GMCs in DYN model.
	Simulated GMCs are identified from spiral area shown in Fig. \ref{fig:Snapshot}c.
	The mass resolution (typical SPH particle mass multiplied times typical number of neighbours) 
	is approximately $2 \times 10^4~\rm M_\odot$, and spatial resolution (FoF method linking length) is $10$ pc.
	Bottom panels: Same as top panels, but for SDW model.
	Simulated GMCs are identified from area shown in Fig. \ref{fig:Snapshot}d.
}	
\label{fig:Scaling}
\end{center}
\end{figure}

Fig. \ref{fig:Scaling} compares the identified GMCs (hereafter, `simulated GMCs') with the observed GMCs 
in terms of the scaling relationships between the cloud masses $M_{\rm cl}$, sizes $R_{\rm cl}$, 
and line-of-sight velocity dispersions $\sigma_v$.
Here, we followed the method used by \citet{Colombo+2014} to calculate $R_{\rm cl}$: 
\begin{eqnarray}
R_{\rm cl} = 1.91 \sqrt{s_{\rm mj}^2 + s_{\rm mn}^2},
\end{eqnarray}
where $s_{\rm mj}$ and $s_{\rm mn}$ are the major and minor radii, respectively, 
and they are evaluated from the moment-of-inertia matrix. $\sigma_v$ is given by:
\begin{eqnarray}
\sigma_v = \sqrt{\sigma_{v,z}^2 + c_s^2},
\end{eqnarray}
where $\sigma_{v,z}$ and $c_s \simeq 1~\rm km~s^{-1}$ are the mass-weighted velocity dispersion 
with respect to the $z$-axis and the mass-weighted sound speed, respectively.
Figs. \ref{fig:Scaling}a and \ref{fig:Scaling}b and Figs. \ref{fig:Scaling}c and \ref{fig:Scaling}d
show that the DYN and SDW models, respectively, reproduced the observed scaling relations well.
Therefore, it can be said that the simulated GMCs in both models follow the observed scaling relations to a reasonable extent.
Furthermore, the simulated scaling relations of the SDW model are similar to those of the DYN model,
suggesting that the statistical properties of GMCs depend primarily on the local environments 
rather than the dynamical nature of the spiral arms.

\subsection{Distributions of GMCs around Spiral Arms}
\label{sec:GMCDistributionAroundSpiral}

\begin{figure*}
\begin{center}
\includegraphics[width=0.95\textwidth]{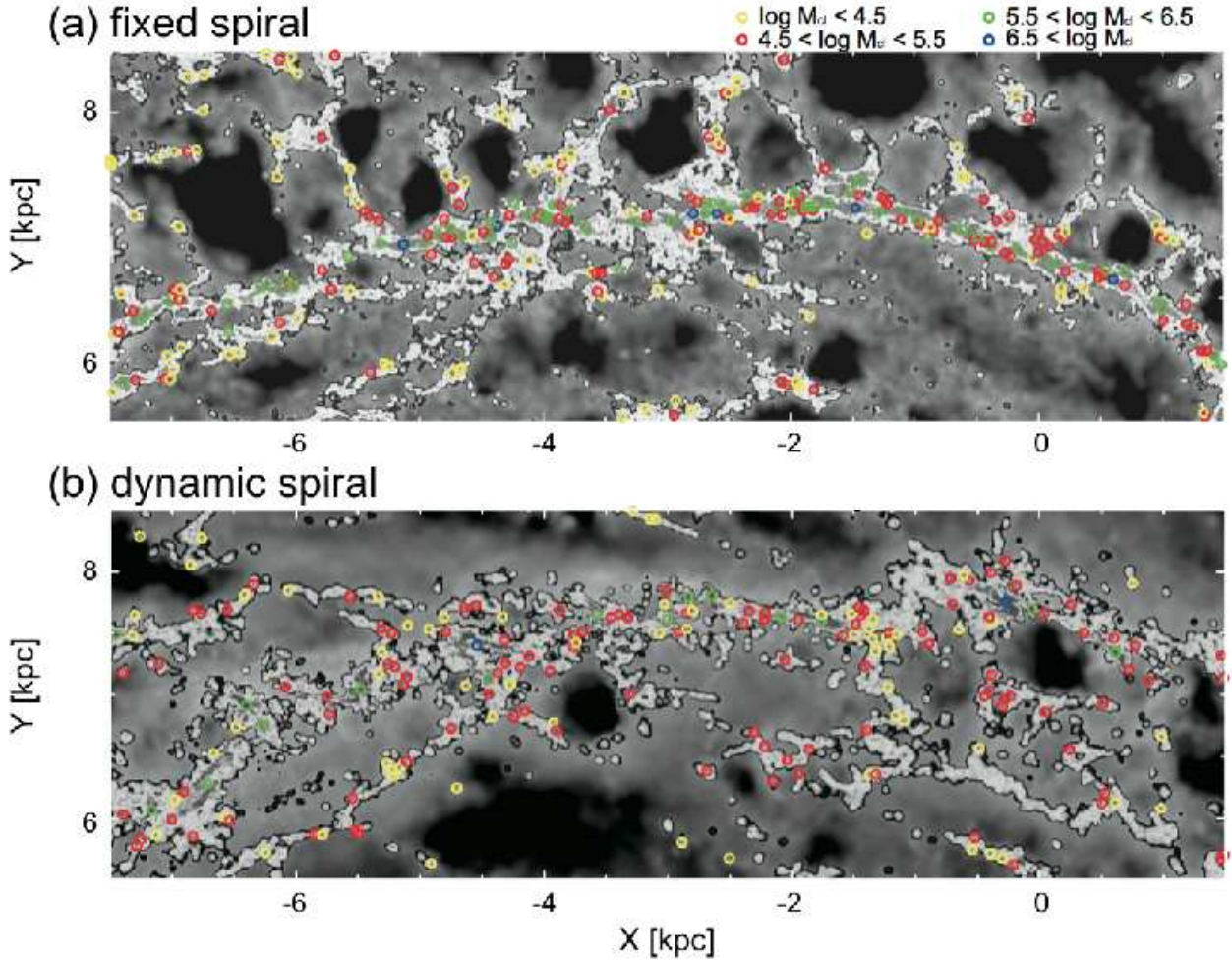}
\caption{
	Distributions of simulated GMCs around the spiral regions in (a) SDW model and (b) DYN model.
	Symbols corresponds to GMC masses. 
	Grey-scale density maps are identical to maps shown in Figs. \ref{fig:Snapshot}(c) and \ref{fig:Snapshot}(d), respectively.
}	
\label{fig:GMCacrossArmMass}
\end{center}
\end{figure*}

We next compare the spatial distributions of the simulated GMCs around the spiral arms in the DYN and SDW models.
In Fig. \ref{fig:GMCacrossArmMass}, the GMCs are overlaid on the density maps of the gas of the two models. 
The colours of the GMCs indicate the GMC masses.
In the SDW model, massive GMCs with $M_{\rm cl} \gtrsim 10^6~\rm M_\odot$ are evident in the spiral arm, 
whereas many smaller GMCs with $M_{\rm cl} \lesssim 10^5~\rm M_\odot$ are apparent in the regions outside of the arm (Fig. \ref{fig:GMCacrossArmMass}a).
This result is consistent with those of observational studies of a spiral galaxy M51 \citep[][]{Koda+2009,Colombo+2014}.
However, the GMC `mass segregation' is also observable in the DYN model (Fig. \ref{fig:GMCacrossArmMass}b).
Furthermore, neither model shows clear mass sequences across the spiral arms.
It is therefore suggested that spatial distributions of GMC masses in spiral galaxies 
do not depend on the dynamical behaviours of their spiral arms.

\begin{figure*}
\begin{center}
\includegraphics[width=0.95\textwidth]{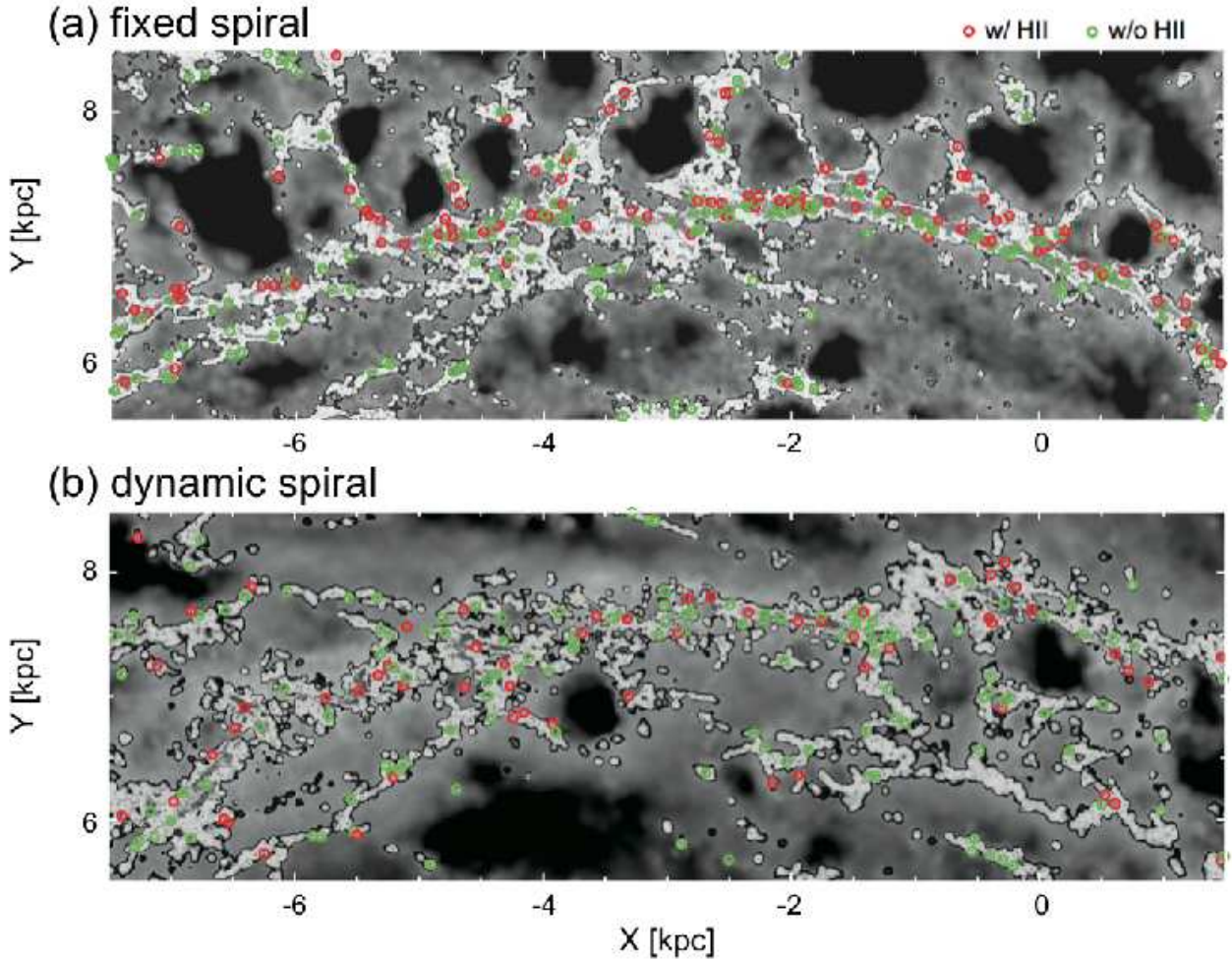}
\caption{
	Same as Fig.\ref{fig:GMCacrossArmMass}, 
	but symbols indicate whether GMCs are associated with $\rm H_{II}$-regions. 
}	
\label{fig:GMCacrossArmHii}
\end{center}
\end{figure*}

On the other hand, Fig. \ref{fig:GMCacrossArmHii} compares the distributions of the GMCs, 
in terms of whether the GMCs are associated with $\rm H_{II}$-regions.
In the SDW model, most of the GMCs with $\rm H_{II}$ regions
are distributed in the spurs and the leading (i.e. downstream) side of the arm over a wide radial range (Fig. \ref{fig:GMCacrossArmHii}a).
This is reasonable since the gas overtakes the spiral arm potential after compressed around the spiral potential trough. 
In contrast, the DYN model yields no clear trend that regarding whether or not 
the GMCs are associated with $\rm H_{II}$ regions (Fig. \ref{fig:GMCacrossArmHii}b).
It should be emphasised here that the evolutionary sequence of the GMCs across the arm is also 
observable in the DYN model {\it locally}, but it does not appear in a wide radial range.

\begin{figure}
\begin{center}
\includegraphics[width=0.45\textwidth]{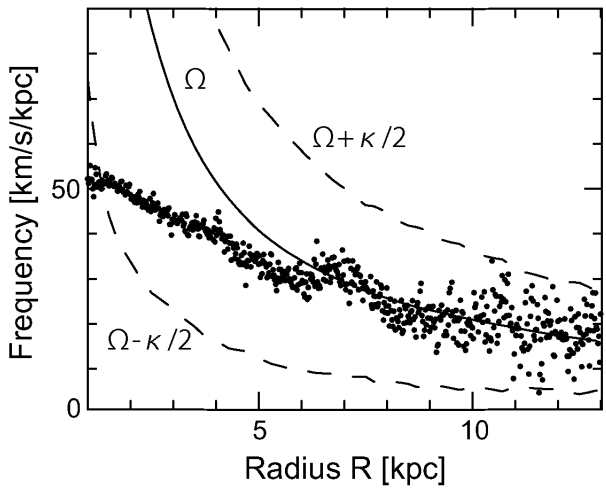}
\bigskip
\caption{
	Radial distribution of angular phase speed $\Omega_{\rm phase}$ of $m=2$ mode in DYN model.
	Time is identical to that in Fig.\ref{fig:Snapshot}.
	Solid and dashed curves indicate rotation angular speed of galaxy ($\Omega$) and $\Omega \pm \kappa/2$ 
	(here $\kappa$ denotes epicyclic frequency), respectively. In $R \gtrsim 6$ kpc regions,  
	spiral arms rotate with $\Omega_{\rm phase} \simeq \Omega$ \citep[see also][]{Baba2015c}.
}
\label{fig:PatternSpeed}
\end{center}
\end{figure}

\begin{figure*}
\begin{center}
\includegraphics[width=0.95\textwidth]{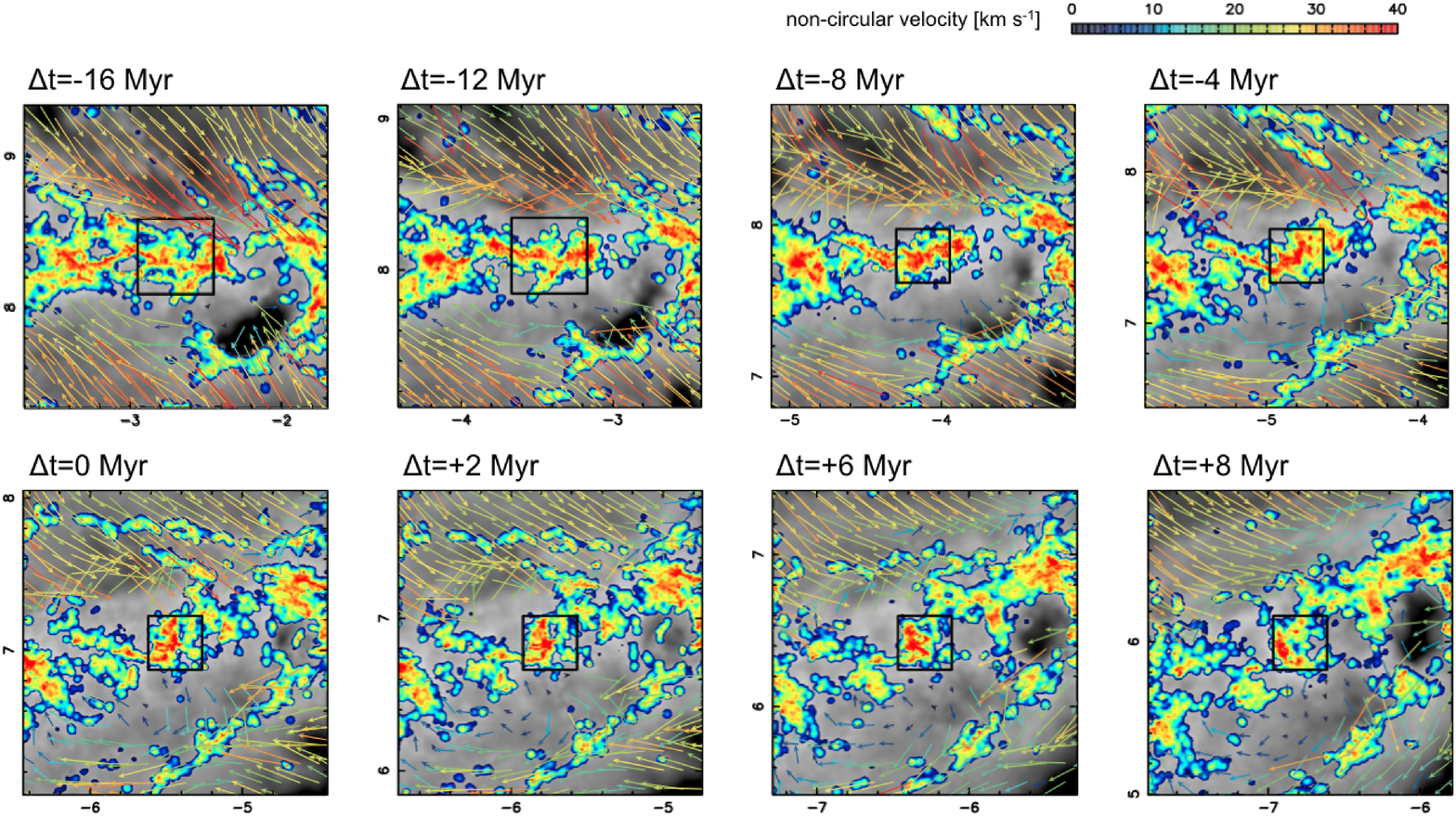}
\caption{
	Evolution of density distributions (in $x$-$y$ plane) around the spiral arm in DYN model.
	The surface densities of the ${\rm H_I}$ (grey) and ${\rm H_2}$ (rainbow) gases are 
	presented at $\Delta t = $ -16, -12, -8, -4, 0, +2,  +6  Myr, and +8 Myr,
	where $\Delta t =0$ corresponds to $t=2.625$ Gyr.
	The non-circular velocity fields are indicated by arrows, 
	the lengths and colours of which are in accordance with the velocities.
	The evolution of the GMC enclosed by the solid square is shown in Fig. \ref{fig:CloudEvolutionZoom}
	and discussed in Section \ref{sec:GMCdetails}.
}	
\label{fig:SpiralEvolutionZoom}
\end{center}
\end{figure*}

The lack of the evolutionary sequences of GMCs in the DYN model originates from the difference 
between the gas flows around spiral arms in the DYN and SDW models.
As predicted by the traditional spiral model, 
in the SDW model, the gas enters into the spiral arm from a single side, is compressed, and then forms stars.
On the other hand, the DYN model has {\it dynamically evolving} spiral arms,
which rotate almost following the galactic rotations well outside of the bar (Fig. \ref{fig:PatternSpeed}). 
In these dynamic spirals, both stellar and gaseous arms are formed simultaneously by `large-scale colliding flows,' 
which originate from the non-linear epicyclic motions of stars and gas 
and subsequently disperse due to the same motions \citep[\cite{Wada+2011}; see also Fig.2 of][]{Baba+2016a}.
In Fig. \ref{fig:SpiralEvolutionZoom}, non-circular velocity fields have been overlaid onto the gas-density maps around 
the regions enclosed by the solid square in Fig. \ref{fig:Snapshot}c.
Non-circular motions of $\sim 30$--$40~\rm km~s^{-1}$ on both sides of the arm are observable, 
along with `large-scale colliding flows', where non-circular velocities on the leading side of the arm 
(the upper side of each panel) indicate flow from the outer to inner region, 
whereas those on the trailing side of the arm (the lower side of each panel) exhibit flow in the opposite direction. 
Thus, no evolutionary sequence of GMCs is observable around the dynamic spirals.

\section{Evolution of GMCs in Dynamic Spirals}
\label{sec:DynamicSpiralResults}

To start, we consider one particular GMC in the DYN model in detail. 
We then address the statistics of GMCs in Sections \ref{sec:Statistics} and \ref{sec:ExtendedVirialAnalysis}.

\subsection{Case Study} 
\label{sec:GMCdetails}

In this subsection, we focus on one simulated GMC in the spiral environment 
(enclosed by the solid square in Fig. \ref{fig:Snapshot}c) and present the details of its formation and evolution.
This simulated GMC had $M_{\rm cl} \simeq 1.3 \times 10^6~\rm M_\odot$, $R_{\rm cl} \simeq 51$ pc, 
and $\sigma_v \simeq 4.8~\rm km~s^{-1}$, which follow the observed scaling relationships well (c.f. Fig. \ref{fig:Scaling}).
This GMC contained about $2000$ SPH particles.

\subsubsection{Hierarchical Collisional Build-Up}

Fig. \ref{fig:CloudEvolutionZoom} shows that the GMC at $\Delta t=0$ grows up via hierarchical agglomeration of 
the several smaller clouds (with a typical mass of $\lesssim 10^5~\rm M_\odot$), 
which starts from 10--20 Myr before the GMC formation.
Focusing on cloud cl-A, which was the main progenitor of this GMC at $\Delta t = 0$, 
this eventful growth can be explained as follows: 
cl-A was formed by the merging of clouds cl-A3 and cl-A4 at $\Delta t  \simeq -9$ Myr,
with the relative velocity for this merger being approximately $10~\rm km~s^{-1}$.
Before this merger, cloud cl-A4 grew from cloud cl-A1 via accumulation of diffuse cl-A2 and diffuse molecular gas.
After cl-A3 merged with cl-A4, cl-A collided with a nearby cloud, cl-B, at $\Delta t \simeq -3$ Myr 
and with cloud cl-C until $\Delta t = -1$ Myr. 
The relative velocities for the cl-A--cl-B and cl-A--cl-C collisions were approximately 
$13~\rm km~s^{-1}$ and $14~\rm km~s^{-1}$, respectively.

The mass evolutions of the progenitor clouds of this GMC are presented in Fig. \ref{fig:CloudHistory}a.
This figure shows that the mergers or collisions occur every $\lesssim 5$ Myr for this GMC. 
Since we analysed only one simulated GMC, we could not quantitatively predict the collision frequency, 
but we note that this collision frequency is comparable to or slightly shorter than those 
estimated from hydrodynamic simulations of fixed spiral potentials \citep{Fujimoto+2014a,Dobbs+2015};
however, it is much shorter than those obtained in no-spiral cases \citep{TaskerTan2009}.
This difference suggests that the GMC collision frequency is increased due to the presence of the spiral arms,
but it may not depend on the dynamics of the spiral arms.

Such hierarchical collisional build-up is driven by the large-scale colliding flows to form a spiral arm, 
as well as a local expanding motion that originates in stellar feedback from a nearby star cluster.
Fig. \ref{fig:CloudEvolutionZoomVelocity} presents the velocity fields 
around the GMC shown in Fig. \ref{fig:CloudEvolutionZoom}. 
Before cl-A forms ($\Delta t$ = -16 Myr and -12 Myr), 
the velocity fields clearly show converging flows, 
which are driven by the large-scale colliding flows.
After cl-A forms, cl-B and cl-C are pushed toward cl-A by an expanding motion associated with a nearby star cluster 
(located at the upper right corner of the panel at $\Delta t = -4$ Myr) and then merge with cl-A.

\subsubsection{Star Cluster Formation and Feedback}

\begin{figure*}
\begin{center}
\includegraphics[width=0.95\textwidth]{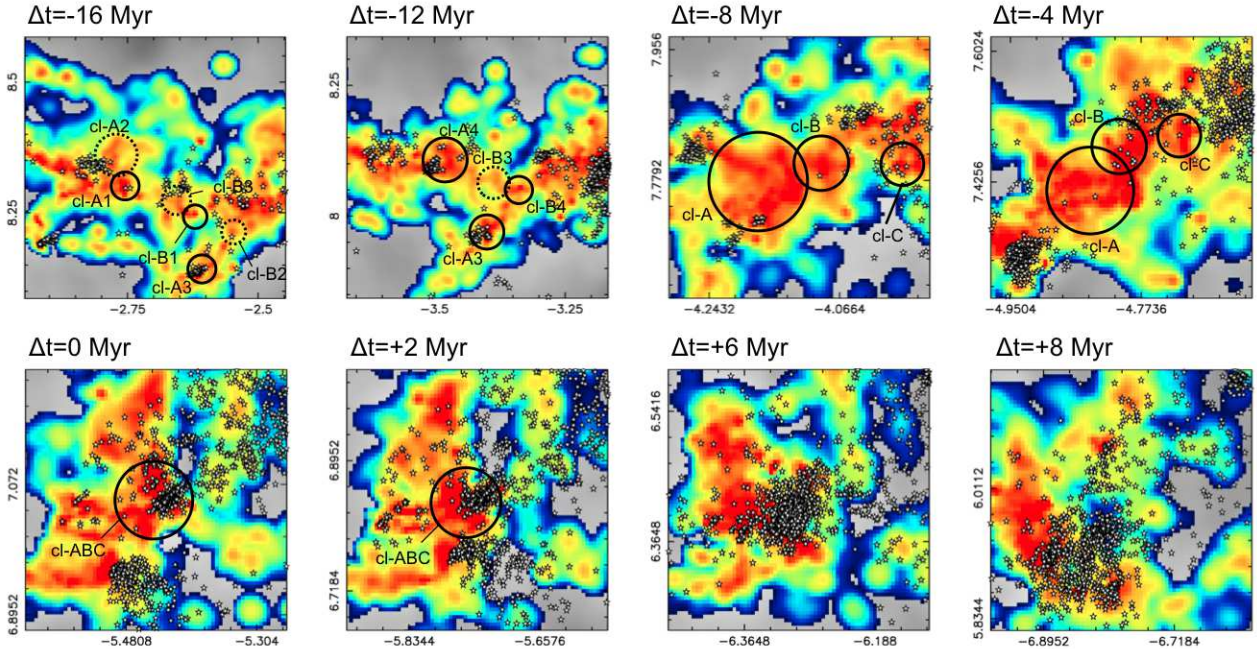}
\caption{
	Evolution of GMC in the spiral environment shown in Fig. \ref{fig:SpiralEvolutionZoom}.	
	The surface densities of the ${\rm H_I}$ (grey) and ${\rm H_2}$ (rainbow) gases are 
	presented at $\Delta t = $ -16, -12, -8, -4, 0, +2, +6, and +8 Myr.
	Circles indicate the progenitor clouds of the GMC (labelled `cl-ABC' at $\Delta t \simeq 0$), and
	the dotted circles are the progenitor clouds with column densities less than $3 \times 10^{21}~\rm H~cm^{-2}$.
	Progenitor clouds were identified based on distribution of SPH particles that make up `cl-ABC'.
	Star particles with OB stars (the open star symbols) have been overlaid on the maps.
	A star cluster starts to form at $\Delta t \simeq -2$ Myr.
}	
\label{fig:CloudEvolutionZoom}
\end{center}
\end{figure*}

\begin{figure*}
\begin{center}
\includegraphics[width=0.7\textwidth]{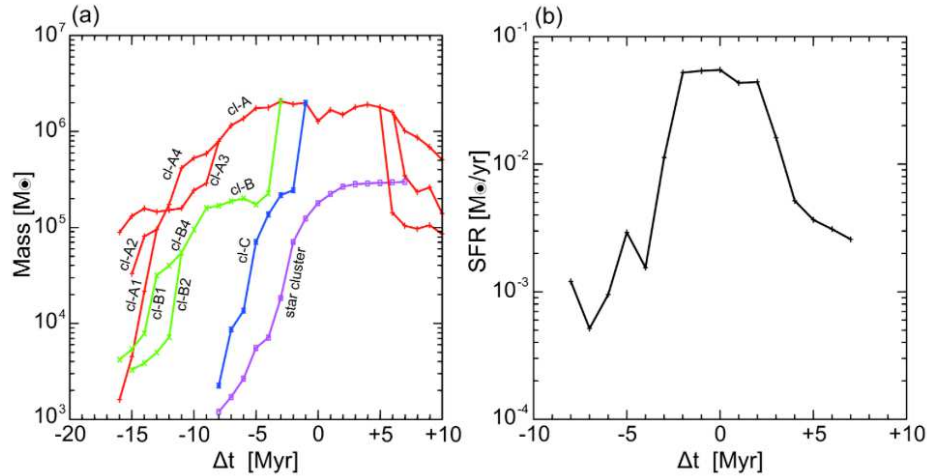}
\bigskip
\caption{
	(a) Evolution of masses of progenitor clouds and star cluster,
	and (b) SFR in the main progenitor cloud shown in Fig. \ref{fig:CloudEvolutionZoom}, cl-A.
}	
\label{fig:CloudHistory}
\end{center}
\end{figure*}

Fig. \ref{fig:CloudEvolutionZoom} shows that a cloud-cloud collision 
between cl-A and cl-B triggers the formation of a star cluster at $\Delta t \simeq -3$ Myr.
Cluster formation due to cloud-cloud collision has been suggested previously, based on recent observations 
\citep{Furukawa+2009,Ohama+2010,Fukui+2014} as well as numerical simulations \citep{HabeOhta1992,InoueFukui2013,Takahira+2014,FujiiPortegiesZwart2016}.
The evolution of the mass fraction for different densities is shown in Fig. \ref{fig:CloudHistoryMassFraction}.
At $\Delta t \simeq -10$ Myr, the dense gas with $n \gtrsim 10^2~\rm cm^{-3}$ occupies 
only approximately 10\% of the mass fraction, although the mass fraction of the highly dense gas
with $n \gtrsim 10^3~\rm cm^{-3}$ increase rapidly at $\Delta t > -5$ Myr. 
Such thermal evolution of the constituent gas is clearly observable in $n$--$T$ plane (Fig. \ref{fig:RawMaterialThermalHistory}).
As shown in Fig. \ref{fig:CloudHistory}b, the SFR in cl-A exhibits a rapid increase after this collision.
These results clearly demonstrate that this collision increases the fraction of highly dense gas, 
and then triggers the formation of a star cluster\footnote{
As suggested by previous simulations of star cluster formation in turbulent molecular clouds \citep[e.g.][]{Fujii2015,FujiiPortegiesZwart2016}, 
growth of a star cluster via hierarchical mergers of smaller star clusters might also be observed in our galactic-scale simulation.
However, we focus on the formation and evolution of GMCs in this paper. 
This topic will be discussed elsewhere.
}.

\begin{figure*}
\begin{center}
\includegraphics[width=0.95\textwidth]{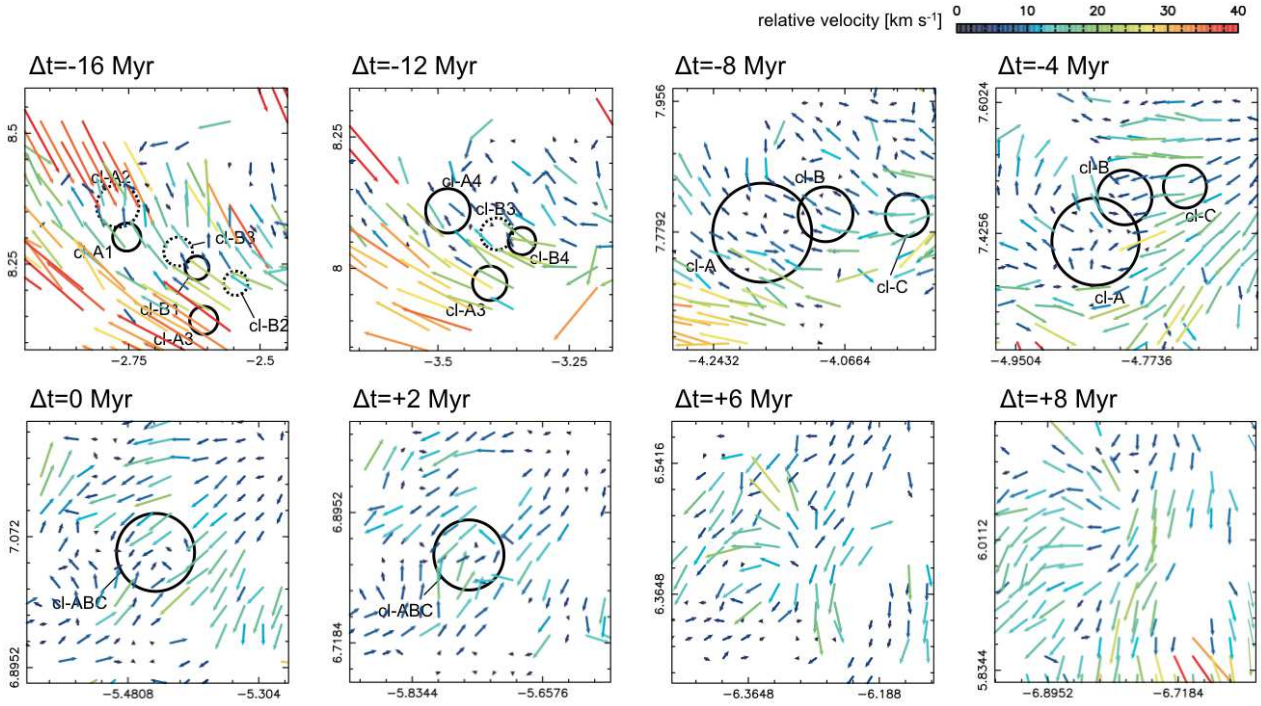}
\bigskip
\caption{
	Evolution of velocity fields around the GMC shown in Fig. \ref{fig:CloudEvolutionZoom}.
	The velocity fields are indicated by the arrows, the lengths and colours of which are in accordance with the relative velocities 
	with respect to the centroid velocity of the constituent gas forming the GMC at $\Delta t = 0$.
}	
\label{fig:CloudEvolutionZoomVelocity}
\end{center}
\end{figure*}

\begin{figure}
\begin{center}
\includegraphics[width=0.4\textwidth]{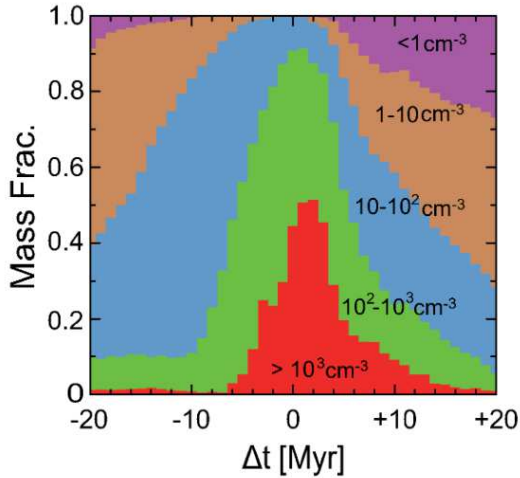}
\bigskip
\caption{
	Evolution of mass fractions of constituent gases of cl-ABC shown in Fig. \ref{fig:CloudEvolutionZoom} with 
	density $n_{\rm H}$ (red: $>10^3~\rm cm^{-3}$, green: $10^2$--$10^3~\rm cm^{-3}$, 
	light blue: $10$--$10^2~\rm cm^{-3}$, orange: 1--10 $\rm cm^{-3}$, and purple: $<1$ $\rm cm^{-3}$).
}	
\label{fig:CloudHistoryMassFraction}
\end{center}
\end{figure}

\begin{figure*}
\begin{center}
\includegraphics[width=0.98\textwidth]{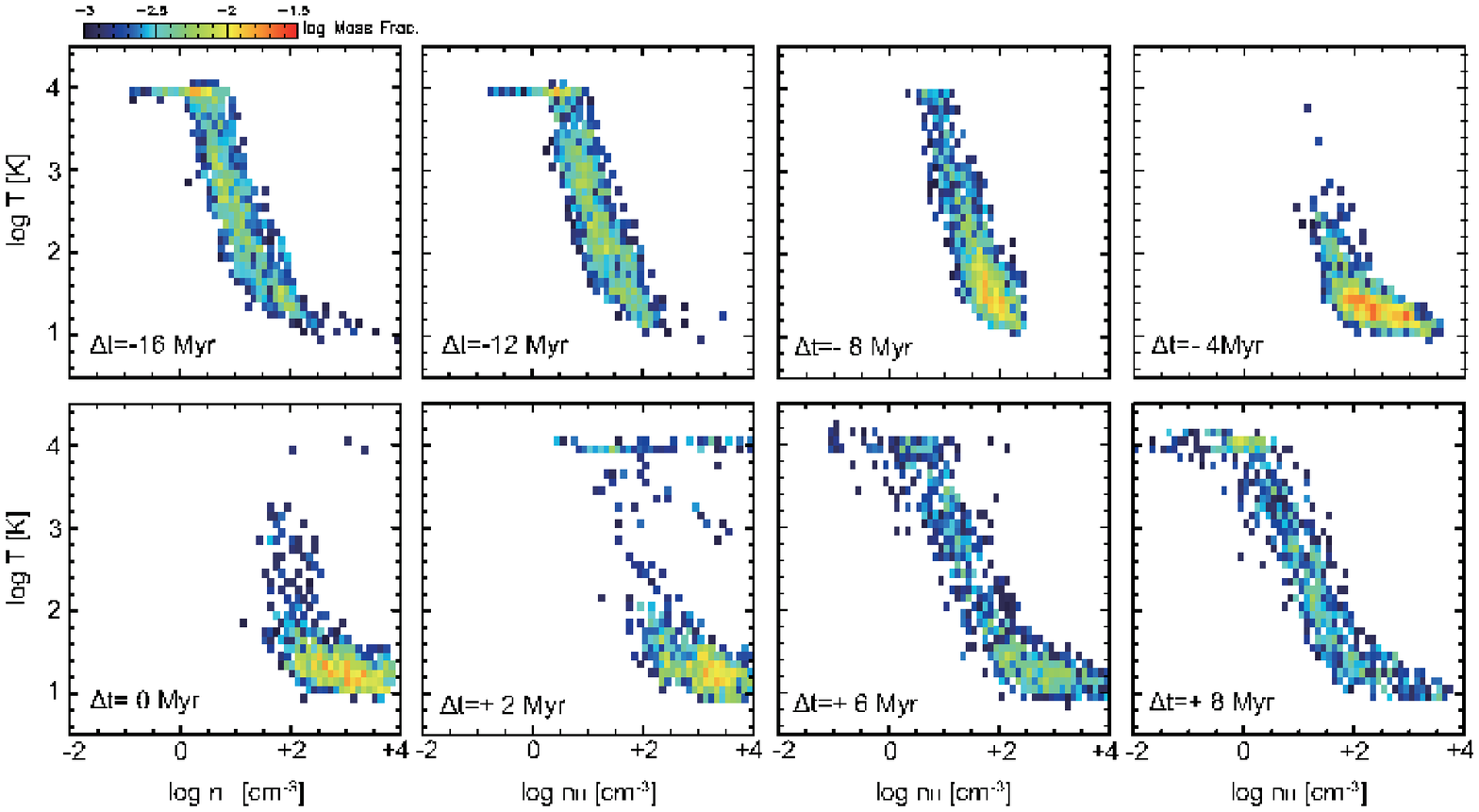}
\bigskip
\caption{
	Thermal evolution of the constituent gas that forms the GMC at $t=2.625$ Gyr shown in Fig. \ref{fig:CloudEvolutionZoom}.
	The colours are coded by differential mass fraction (on a logarithmic scale).
}	
\label{fig:RawMaterialThermalHistory}
\end{center}
\end{figure*}

The star cluster, which is formed by the cloud-cloud collision between cl-A and cl-B,
leads to destruction of the GMC (i.e. cl-ABC) via ${\rm H_{II}}$-region feedback at $\Delta t \gtrsim 2$ Myr.
In fact, it is evident that a part of the constituent gas forms 
an ${\rm  H_{II}}$ region at $\Delta t \simeq 2$ Myr
(i.e. $n \gtrsim 10~\rm cm^{-3}$ and $T \simeq 10^4$ K; see Fig. \ref{fig:RawMaterialThermalHistory}).
Simultaneously, as shown in Fig. \ref{fig:CloudHistory}b, the SFR in cl-A exhibits a sudden decrease at $\Delta t > 2$ Myr
because of the dispersion and consumption of the dense gas of $n \gtrsim 10^3~\rm cm^{-3}$
(see also Fig. \ref{fig:CloudHistoryMassFraction}).
The expanding motion around the star cluster can be clearly seen at $\Delta t$ = 6 Myr 
in Fig. \ref{fig:CloudEvolutionZoomVelocity}.
Note that the ${\rm H_{II}}$-region feedback does not completely destroy the molecular gas in the cl-ABC,
but rather breaks up the cl-ABC into other smaller clouds of $\sim 10^5~\rm M_\odot$
at $\Delta t \simeq 5$ Myr and  6 Myr (see Fig. \ref{fig:CloudHistory}a).

Because determining GMC lifetimes is highly non-trivial, we here present two GMC lifetime estimates 
that were obtained using two different methods.
First, by defining the lifetime of this GMC as the period between the last major merger ($\Delta t \simeq -1.5$ Myr) 
and the destruction time ($\Delta t \simeq 6$ Myr), we obtained a lifetime of approximately 7 -- 8 Myr for this GMC. 
Secondly, by defining the lifetime as being from the time when this GMC reached half of its maximum mass
($\Delta t \simeq -10$ Myr) to the destruction time, a lifetime of approximately 15 Myr was obtained.
In this sense, we can say that the lifetime of this GMC is approximately 10 -- 15 Myr.
This lifetime is comparable to the GMC's free-fall time and is consistent with the values inferred from 
recent observational studies of molecular clouds in the Large Magellanic Cloud \citep{Kawamura+2009} and 
the distributions of young star clusters around spiral arms in nearby spiral galaxies \citep{Elmegreen2007}.
It is also noted that the lifetime of this simulated GMC is in reasonable agreement with the values 
suggested by previous numerical studies \citep{DobbsPringle2013,Fujimoto+2014a}.

\subsection{Statistics of Dynamical State of GMCs}
\label{sec:Statistics}

In Section \ref{sec:GMCdetails}, we focused on a simulated GMC and showed that the GMC underwent 
eventful evolution including hierarchical agglomerations of smaller clouds and collision-induced star formation 
and subsequent destruction via stellar feedback. These results suggest that GMCs are dynamic and transient structures.
In this subsection, we focus on statistical properties of the dynamical states of the simulated GMCs.

To investigate the dynamical properties of the simulated GMCs, 
we introduce a quantity $\langle \nabla\cdot{\bf v} \rangle$, which is given by
\begin{eqnarray}
 \langle \nabla\cdot{\bf v} \rangle = \frac{1}{V} \sum_{i} \frac{m_i}{\rho_i} \nabla\cdot{\bf v}_i,
 \label{eq:div}
\end{eqnarray}
where ${\bf v}_i$, $m_i$, and $\rho_i$ are the velocity, mass, and mass density, respectively, 
of the $i$-th constituent SPH particle of a GMC, and the total volume $V = \sum_{i} m_i/\rho_i$.
Equation (\ref{eq:div}) represents a volume-weighted average of $\nabla\cdot{\bf v}$
over the constituent SPH particles of a GMC and measures whether a GMC is {\it globally} collapsing or expanding:
$\langle \nabla\cdot{\bf v} \rangle <0$ indicates that a GMC is globally collapsing, 
while an expanding GMC has $\langle \nabla\cdot{\bf v} \rangle >0$.

\begin{figure*}
\begin{center}
\includegraphics[width=0.95\textwidth]{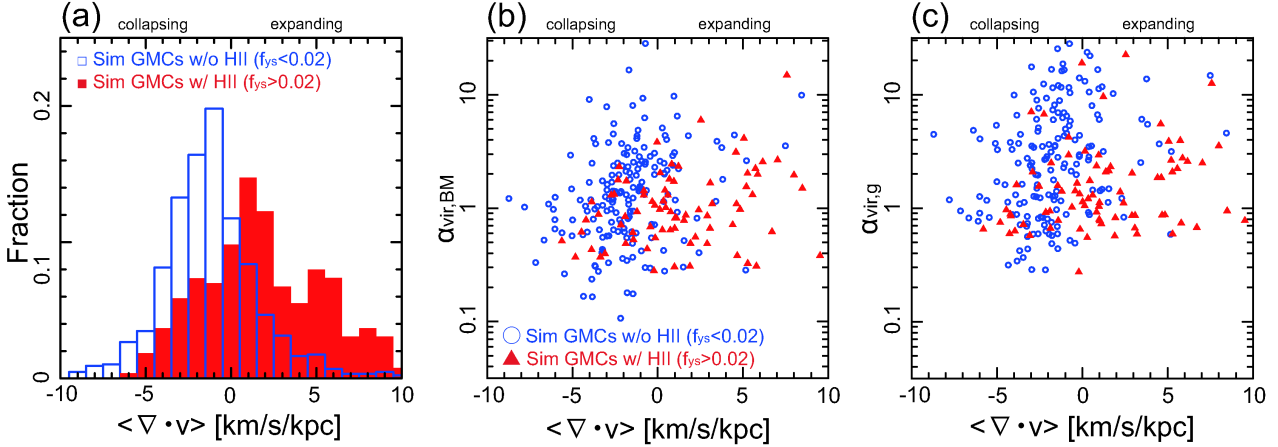}
\bigskip
\caption{
	Statistical properties of the simulated GMCs shown in Fig. \ref{fig:Scaling}.
 	(a) Probability distribution functions of $\langle \nabla\cdot{\bf v} \rangle$ of the simulated GMCs
	with and without ${\rm H_{II}}$ regions.
 	(b) Distributions of the simulated GMCs on $\langle \nabla\cdot{\bf v} \rangle$--$\alpha_{\rm vir,BM}$ plane.
	(c) Same as panel (b), but for $\alpha_{\rm vir,g} \equiv 2E_{\rm kin}/|E_{\rm g}|$.
}	
\label{fig:Statistics}
\end{center}
\end{figure*}

In Fig. \ref{fig:Statistics}a, probability distributions of the $\langle \nabla\cdot{\bf v} \rangle$ 
for the simulated GMCs are presented. To see the effect of the $\rm H_{II}$-region feedback, 
we divided the GMCs into subsamples according to the mass ratio of 
young stellar particles within the GMC to the GMC mass, i.e. $f_{\rm ys} \equiv M_{\rm ys}/M_{\rm cl}$, 
where $M_{\rm ys}$ is the total mass of stellar particles containing OB stars within the radius of the GMC.
These distributions clearly show that most of the simulated GMCs {\it without} ${\rm H_{II}}$ regions (i.e. $f_{\rm ys} < 0.02$) 
are collapsing\footnote{
In general, the word `collapse' is used to refer to `gravitational collapse'.
However, in this paper, 
we use this term purely to indicate `inward motions'.
}. 
In contrast, more than half of the simulated GMCs {\it with} ${\rm H_{II}}$ regions (i.e. $f_{\rm ys} \geq 0.02$)
are expanding ($\nabla\cdot{\bf v} > 0$).

These results support the `dynamic' picture rather than the traditional virialized picture.
In the traditional picture, GMCs are thought to be {\it virialized} structures \citep[][]{ZuckermanEvans1974,Heyer+2009}, 
and the internal supersonic turbulence of a GMC is thought to support the cloud against global gravitational collapse and 
to regulate star formation with a low efficiency \citep[i.e. turbulent-regulated star formation; e.g. ][]{KrumholzMcKee2005,KrumholzTan2007}.
Continuous energy injections are required to sustain this picture, although these energy sources are not well understood.
Alternatively, the dynamic picture has been developed in which GMCs undergo collapse 
rather than existing in virial equilibrium \citep{GoldreichKwan1974,Ballesteros-Paredes+2011b} 
and are destroyed by stellar feedback from internal stars \citep[e.g.][]{Elmegreen2007,Murray2011}. 
In this picture, stellar feedback prevents star formation before all the gas turns into stars (i.e. feedback-regulated star formation).
It is worth emphasizing that our results suggest not only that the GMCs are collapsing, 
but also that they suffer from collisional build-up and collision-induced star formation.

The `dynamic' picture of GMCs does not contradict previous observations. Fig. \ref{fig:Statistics}b shows 
the distribution of the simulated GMCs on the $\langle \nabla\cdot{\bf v} \rangle$--$\alpha_{\rm vir,BM}$ plane.
Here, $\alpha_{\rm vir,BM}$ is the virial parameter of a GMC and is defined as
\begin{eqnarray}
\alpha_{\rm vir,BM} \equiv \frac{5 \sigma_v^2 R_{\rm cl}}{GM_{\rm cl}}.
\end{eqnarray}
This quantity can be evaluated from the observable quantities of GMCs 
\citep[i.e. $M_{\rm cl}$, $R_{\rm cl}$, and $\sigma_v$;][]{BertoldiMcKee1992}\footnote{
This definition does not account for galactic tidal or extra pressure terms.
See Section \ref{sec:ExtendedVirialAnalysis} for the effects of these terms.
}. 
The simulated GMCs typically have $\alpha_{\rm vir,BM} \gtrsim 1$, 
which is consistent with previously reported observed values \citep[e.g.][]{Heyer+2009}.
However, no clear correlation is evident between $\langle \nabla\cdot{\bf v} \rangle$ and $\alpha_{\rm vir,BM}$,
suggesting that $\alpha_{\rm vir,BM}$ is not a good indicator of the dynamical state of a GMC. 
$\alpha_{\rm vir,BM}$ describes the balance between the gravitational and kinetic energies in a cloud, 
but it does not indicate whether the cloud is expanding or contracting.
In fact, if a cloud is converging, i.e. moving in the same direction as the force of gravity 
but not with a random turbulent motion, $\alpha_{\rm vir,BM}$ will be large.
In other words, neither does $\alpha_{\rm vir,BM} > 1$ strictly corresponds to expansion, 
nor does $\alpha_{\rm vir,BM} < 1$ strictly correspond to contraction \citep{Ballesteros-Paredes2006}.
Thus, the observation of GMCs having $\alpha_{\rm vir,BM} \simeq 2$ cannot be considered
evidence in support of the fact that these are in the virial equilibrium state.

We here note that $\alpha_{\rm vir,BM}$ does not necessarily apply to elongated GMCs, 
since the above definition of $\alpha_{\rm vir,BM}$ is based on 
the assumption that a GMC is a uniform sphere \citep{BertoldiMcKee1992}.
To check the validity of this assumption, 
we introduced a generalized viral equation by considering a GMC moving in a galactic potential 
(see Appendix \ref{sec:VirialTheoremInGalaxies} for details) as follows:
\begin{eqnarray}
 \frac{1}{2}\frac{{\rm D}^2 I}{{\rm D}t^2} = 2 E_{\rm kin} - \mathcal{T}_{\rm p} + E_{\rm g} + W_{\rm tidal}, 
 \label{eq:GeneralVirialEquation}
\end{eqnarray}
where ${\rm D}/{\rm D}t$ denotes the Lagrangian time derivative, $I$ is the moment of inertia of the cloud; 
$E_{\rm kin}$ and $E_{\rm g}$ are the internal kinetic energy and self-gravitational energy of the cloud, respectively; 
$\mathcal{T}_{\rm p}$ and $W_{\rm tidal}$ are the works by an external pressure and a gravitational force 
due to the mass outside the cloud 
(i.e. galactic tides including the centrifugal/Coriolis forces on the cloud), respectively\footnote{
We ignored the thermal energy of the cloud because it is much smaller than the other terms.}. 
By neglecting the terms for galactic tides and external pressure (see also Section \ref{sec:ExtendedVirialAnalysis}), 
the following viral parameter was defined:
\begin{eqnarray}
 \alpha_{\rm vir,g} \equiv - \frac{2 E_{\rm kin}}{E_{\rm g}}, 
\end{eqnarray}
such that a cloud has ${\rm D}^2I/{\rm D}t^2 > 0$ if $ \alpha_{\rm vir,g} >0$.
This definition is independent of the density distribution inside the cloud.
Fig. \ref{fig:Statistics}c shows the distribution of the simulated GMCs 
on the $\langle \nabla\cdot{\bf v} \rangle$--$\alpha_{\rm vir,g}$ plane.
It is evident that $\alpha_{\rm vir,g}$ is systematically larger than $\alpha_{\rm vir,BM}$, 
but there is no clear correlation between $\langle \nabla\cdot{\bf v} \rangle$ and $\alpha_{\rm vir,g}$.
Thus, the above conclusion does not depend on the definition of the virial parameter, 
$\alpha_{\rm vir,BM}$ or  $\alpha_{\rm vir,g}$.

\subsection{Effects of Galactic Tides and External Pressures}
\label{sec:ExtendedVirialAnalysis}

In Section \ref{sec:GMCdetails}, 
we suggested that the formation and evolution of GMCs are significantly affected by environment factors.
We here focus on the environmental effects, particularly galactic tides and external pressures, 
on the dynamical states of GMCs.
To analyse these environmental effects, we used the generalized viral equation (Eq. (\ref{eq:GeneralVirialEquation})) 
and evaluated the probability distribution functions (PDFs) of $W_{\rm tidal,xy}/|E_{\rm g}|$, 
$W_{\rm tidal,z}/|E_{\rm g}|$ and $\mathcal{T}_{\rm p}/|E_{\rm g}|$ for the simulated GMCs, 
as these are rough indicators of the dynamical states of GMCs 
(see Appendix \ref{sec:App2} for the definitions of these quantities). 
Specifically,  $W_{\rm tidal}/|E_{\rm g}| \ll -1$ or $\mathcal{T}_{\rm p}/|E_{\rm g}| \gg 1$, 
then ${\rm D}^2I/{\rm D}t^2 \ll 1$, which implies that the GMC is in a collapsing state\footnote{
As noted by \citet{Ballesteros-Paredes2006}, the sign of ${\rm D}^2I/{\rm D}t^2$ does {\it not} 
determine whether the cloud is contracting or expanding.
}.

\begin{figure*}
\begin{center}
\includegraphics[width=0.95\textwidth]{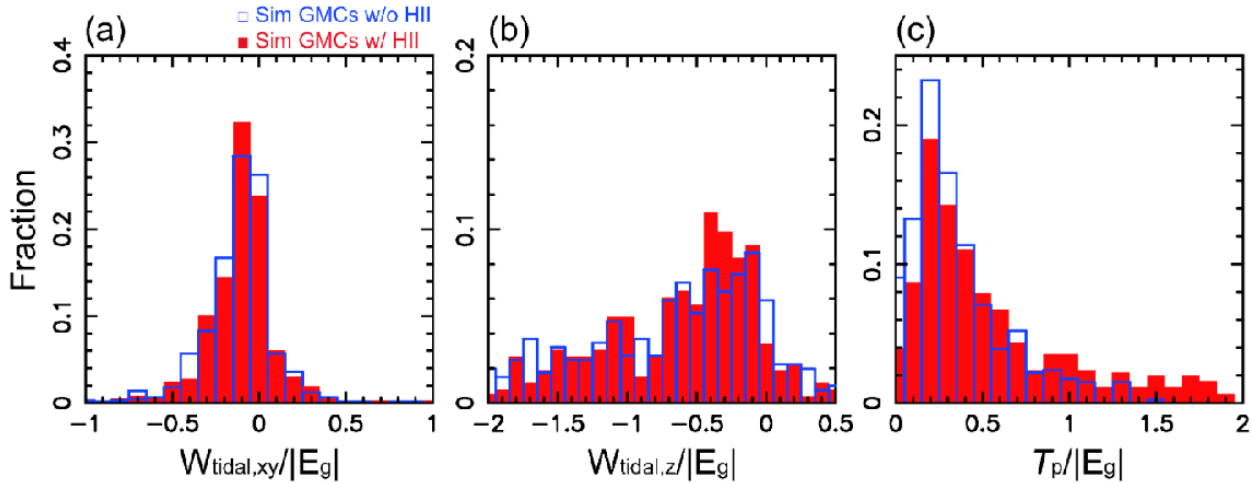}
\bigskip
\caption{
	PDFs of (a) $W_{\rm tidal,x,y}/|E_{\rm g}|$, (b) $W_{\rm tidal,z}/|E_{\rm g}|$, 
	and (c) $\mathcal{T}_{\rm p}/|E_{\rm g}|$ of the simulated GMCs shown in Fig. \ref{fig:Scaling}.
	Definitions of these quantities are given in Appendix \ref{sec:App2}.
}	
\label{fig:StatisticsExtendedVirial}
\end{center}
\end{figure*}

Figs. \ref{fig:StatisticsExtendedVirial}a and \ref{fig:StatisticsExtendedVirial}b show 
the PDFs of $W_{\rm tidal,xy}/|E_{\rm g}|$ and $W_{\rm tidal,z}/|E_{\rm g}|$, 
which are measures of the contributions of the galactic tides on and toward the galactic plane, respectively, 
to the dynamical states of the GMCs. The PDFs of $W_{\rm tidal,xy}/|E_{\rm g}|$ show nearly symmetric distributions 
with peaks at $W_{\rm tidal,xy}/|E_{\rm g}| \simeq -0.1$ and weak tails toward the negative values. 
These results indicate that the galactic tides on the galactic plane do not significantly contribute to GMC collapse.
In contrast, the PDFs of $W_{\rm tidal,z}/|E_{\rm g}|$ have peaks around -0.4 
and long tails toward the negative values, some of which reach $W_{\rm tidal,z}/|E_{\rm g}| < -1$,
whereas there are few GMCs with $W_{\rm tidal,z}/|E_{\rm g}|>0$.
Thus, the contribution of galactic tides toward the galactic plane to the collapse of the GMC 
is half of ($|W_{\rm tidal,z}/E_{\rm g}| \simeq 0.5$) or greater than ($|W_{\rm tidal,z}/E_{\rm g}| > 1$) 
that from the self-gravity.
In addition, the above result does not depend on whether the GMCs have $\rm H_{II}$ regions.

\begin{figure}
\begin{center}
\includegraphics[width=0.45\textwidth]{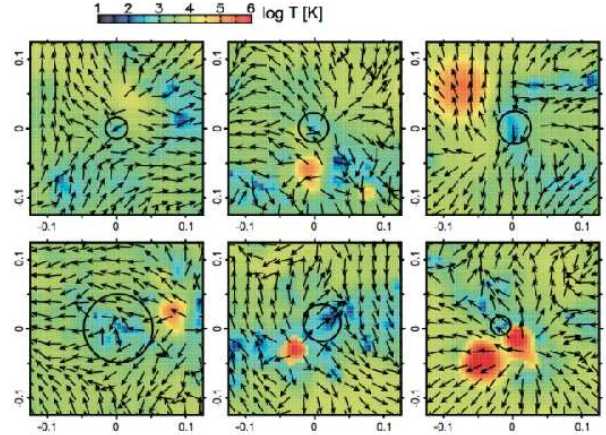}
\bigskip
\caption{
	Zoomed-in temperature maps around GMCs with $\mathcal{T}_{\rm p}/|E_{\rm g}| > 1$.
	The circles indicate the GMC, and the arrows indicate the relative velocities 
	(mass-weighted) with respect to the centroid velocity of the GMC.
	The units of $x$- and $y$- axes are kpc.
}	
\label{fig:SNdrivenSF}
\end{center}
\end{figure}

The external pressure also contributes to the pressure confinement and/or collapse of the simulated GMCs.
Fig. \ref{fig:StatisticsExtendedVirial}c shows that the PDFs of $\mathcal{T}_{\rm p}/|E_{\rm g}|$ have 
peaks around $0.2$ and clear tails at $\mathcal{T}_{\rm p}/|E_{\rm g}| > 1$.
Interestingly, the PDF for the GMCs with the $\rm H_{II}$ regions exhibits a tail at $\mathcal{T}_{\rm p}/|E_{\rm g}| > 1$ 
that is clearer than that for the GMCs without the $\rm H_{II}$ regions. 
Specifically, the typical GMCs (i.e. $\mathcal{T}_{\rm p}/|E_{\rm g}| \sim 0.1$) have $\mathcal{T}_{\rm p} \sim 10^3~\rm K~cm^{-3}$,  
whereas the GMCs with high-$\mathcal{T}_{\rm p}/|E_{\rm g}|$ values have $\mathcal{T}_{\rm p} \gtrsim 10^4~\rm K~cm^{-3}$.
This finding suggests that the high-$\mathcal{T}_{\rm p}/|E_{\rm g}|$ GMCs are confined or pressed by high external pressure.
Fig. \ref{fig:SNdrivenSF} presents zoomed-in maps around the GMCs with $\mathcal{T}_{\rm p}/|E_{\rm g}| > 1$. 
As is evident from this figure, most of the high-$\mathcal{T}_{\rm p}/|E_{\rm g}|$ GMCs 
are located near the shells driven by SN hot bubbles, suggesting that 
star formation activities in high-$\mathcal{T}_{\rm p}/|E_{\rm g}|$ GMCs might be enhanced 
by nearby SN explosions.

\section{Evolution of GMCs in Fixed Spirals}
\label{sec:FixedSpiralResults}

To highlight the characteristics of GMC evolution in dynamic spirals, 
we also discuss GMC evolution in the SDW model. 
However, since the behaviours of GMCs in the SDW model are qualitatively similar to 
those observed using previous hydrodynamic simulations of fixed spiral potentials \citep[e.g.][]{DobbsPringle2013},
we do not present the details of GMC evolution in the SDW model here.

\begin{figure*}
\begin{center}
\includegraphics[width=0.95\textwidth]{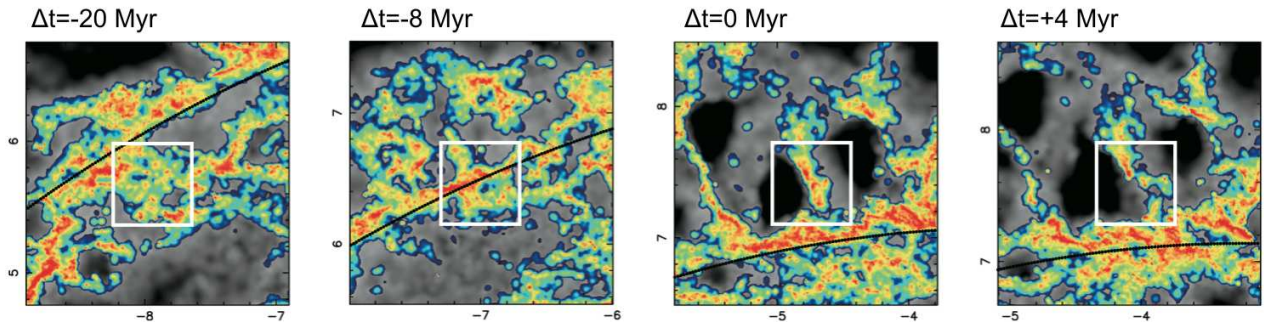}
\bigskip
\caption{
	Evolution of GMC (enclosed by the solid square) in the spiral arm of the SDW model.
	The surface densities of the ${\rm H_I}$ (grey) and ${\rm H_2}$ (rainbow) gases are 
	presented at $\Delta t = $ -16, -8, 0,  and +4  Myr,
	where $\Delta t =0$ corresponds to $t=340$ Myr.
	The black curves indicate the locations of the spiral potential minima.
}	
\label{fig:ShearedGMCinSTEDY}
\end{center}
\end{figure*}

Fig. \ref{fig:ShearedGMCinSTEDY} shows an example of GMC evolution in the SDW model.
This cloud forms the spur enclosed by the solid square in Fig.\ref{fig:Snapshot}d.
It is evident that gas enters the spiral arm potential from the trailing side ($\Delta t = $ -20 Myr)
and is compressed by a convergent flow between the entering gas and the pre-existing gas in the arm ($\Delta t = $ -8 Myr).
During the investigated time, the GMC grows via the collisions of many smaller clouds. 
After passing through the arm, the GMC disperses via shear and results in the formation of a spur ($\Delta t > $ 0 Myr).
Similar behaviours can be observed in other GMCs in the SDW model, 
suggesting that GMCs generally form primarily via the agglomeration of many smaller clouds 
and then disperse when they leave the spiral arms.

This destruction process is the most notable difference between the GMC evolution sequences in the DYN and SDW models,
and it causes the `single-side spurs' that occur in the SDW model (see Section \ref{sec:Spurs}).
Such `shear-driven' GMC evolution is usually yielded by hydrodynamic simulations of fixed spiral potentials 
\citep[][]{WadaKoda2004,DobbsBonnell2006,Wada2008,Dobbs2008,DobbsPringle2013}
and is suggested by the  results of observational studies of the grand-design spiral galaxy M51 \citep{Koda+2009,Miyamoto+2014}.

It should be noted that the gaseous arms do not remain at the same positions; instead, 
they move back and forth between downstream and upstream of 
the spiral potential minima \citep[see also][]{Wada2008}.
In fact, the gaseous arms are located upstream from the spiral arms at $\Delta t = $ -20 Myr, 
on the spiral arms at $\Delta t = 0$ Myr, and downstream from the spiral arms at $\Delta t = $ 4 Myr.
Such oscillating or stochastic behaviours drive the complicated flows within spiral arms
and result in the destruction of the evolutionary sequences of GMCs across spiral arms, 
which would be expected simply based on the traditional spiral model.

\section{Summary}
\label{sec:Summary}

We performed a three-dimensional $N$-body/SPH simulation of a barred spiral galaxy at parsec-scale resolution 
and investigated the dynamical states, formation, and evolution of GMCs in `dynamic' spiral arms.
Our main findings and suggestions are as follows.
\begin{enumerate}
\item The simulated GMCs did not show systematic evolutionary sequences 
(in their masses and star formation activities) across dynamic spiral arms, 
in contrast to the expectation from traditional quasi-stationary density wave 
plus galactic shock theory \citep[e.g.][]{Fujimoto1968,Roberts1969,Shu+1972}.
Investigations of whether GMCs show evolutionary sequences across spiral arms in {\it a wide radial range}
will be a possible means of discriminating the origins of spiral arms. 
These studies will require high spatial resolution and wide-field mapping of molecular gas in spiral galaxies 
with equipment such as the Atacama Large Millimeter/submillimeter Array (ALMA).

\item The simulated GMCs were highly dynamic and exhibited eventful lives involving collisional build-up, 
collision-induced star formation, and destruction via stellar feedback 
before all of the gas within the GMCs were transformed into stars.
These findings are consistent with recent galactic-scale hydrodynamic simulations 
\citep[e.g.][]{Dobbs+2011b,DobbsPringle2013,Hopkins+2011,Hopkins+2012}  
and observations \citep[e.g.][]{Kawamura+2009,Murray2011,Fukui+2014}.
The collisional build-up was driven by large-scale colliding flows associated 
with the spiral arm formation \citep{Wada+2011,Baba+2016a}, as well as by nearby SN explosions. 

\item Although the simulated GMCs were observed to be {\it collapsing} rather than in virial equilibrium, 
they followed the observed scaling relationships well. 
Thus, our results support the dynamic picture of GMCs \citep[e.g.][]{GoldreichKwan1974} rather than 
the traditional equilibrium picture \citep[e.g.][]{ZuckermanEvans1974}.
The global collapse of molecular clouds is supported by recent observations 
\citep[e.g.][]{Schneider+2010,Peretto+2013,Ragan+2015}.

\item The scaling relationships of the GMCs did not originate from the equilibrium state;
instead, these could be explained by considering the collapsing state.
Furthermore, this finding implies that a virial parameter is not a good indicator of 
the equilibrium state of a GMC \citep{Ballesteros-Paredes2006}.
In contrast, through hydrodynamic simulations, 
\citet{Dobbs+2011a} demonstrated that GMCs are predominantly gravitationally unbound objects, 
although their arguments were based on the virial parameter.
In fact, even if GMCs have $\alpha_{\rm vid,BM} > 1$, the GMCs would indicate $\langle \nabla\cdot{\bf v} \rangle < 0$, 
suggesting that the GMCs are collapsing. 

\item The effects of galactic tides and external pressure on the self-gravitational energy of 
the simulated GMCs were found to be non-negligible.
This result suggests that both galactic tides (in particular, the components toward the galactic plane) 
and external pressure contributes to GMC collapse.
Thus, our model suggests that a part of star formation activity in spiral arms is induced 
through the combined effects of increased rates of cloud-cloud collisions and compressions by SN explosions.

\end{enumerate}

According to our numerical resolutions and feedback models, 
GMCs are likely destroyed by stellar feedback (in particular, $\rm H_{II}$-region feedback).
Nevertheless, some previous studies suggested 
that $\rm H_{II}$-region feedback only weekly affects GMC destruction.
\citet{Renaud+2013} implemented the $\rm H_{II}$-region feedback 
using a Stromgren volume approach, which is similar to ours (see Section \ref{sec:SF}), 
into adaptive-mesh refinement (AMR) simulations of a Milky Way-like galaxy.
They argued that $\rm H_{II}$-region heating is not expected to destroy clumps 
but is likely to modify their inner structures and the ongoing star formation.
More recently, \citet{MacLachlan+2015} applied the post-process calculation of 
radiative transfer to a time series of SPH simulations of a spiral galaxy,
and then suggested that the $\rm H_{II}$-region feedback may play only a minor role in the regulation of star formation.

One possible reason for the discrepancy between the results of our study and 
those of the previous studies is the different numerical resolutions.
If the stellar feedback is input into the neighbouring particles or cells, 
the use of a coarse resolutions could cause a large amount of gas to be affected.
However, local-scale hydrodynamic simulations of individual GMCs, 
which were based on a simple ray-tracing algorithm and a Stromgren volume technique, 
showed that GMCs with masses up to $\sim 10^5~\rm M_\odot$ 
could be readily destroyed by the $\rm H_{II}$-region feedback \citep{Dale+2012,Dale+2013}.
Therefore, the effect of stellar feedback could depend on the numerical resolutions, 
hydrodynamic schemes (AMR or SPH), and how the $\rm H_{II}$-region feedback is 
introduced into the simulations \citep[see a review by][and the references therein]{Dale2015review}.
To reach a conclusion, a more sophisticated treatment of stellar feedback within a GMC, as well as higher resolutions, are required.

Overall, the findings of this study imply that  {\it both} galactic structures and local stellar feedback 
are important factors in GMC formation and evolution.
It is worth re-emphasizing that the dynamical effects of galactic tides and external pressure on 
GMCs are not negligible.
Thus, to understand formation and evolution of GMCs, as well as star formation in galaxies, 
more sophisticated treatment of stellar feedback within GMCs should be coupled with galactic-scale simulations.
To explore the dynamical interactions between the ISM and time-dependent stellar structures 
such as spiral arms and bars, along with their environmental dependence, 
we will present detailed studies of these subjects using  self-consistent simulations 
(with parsec-scale resolution) of barred spiral galaxies in forthcoming papers (Baba et al. in preparation).

\section*{Acknowledgements}

We are grateful to the anonymous referee for a number of constructive comments and careful reading. 
We thank Keiichi Wada, Jin Koda and Michiko S. Fujii for helpful comments. 
Calculations, numerical analyses and visualization were carried out on Cray XC30
and computers at Center for Computational Astrophysics, National Astronomical Observatory of Japan.  
JB was supported by HPCI Strategic Program Field 5 `The origin of matter and the universe'
and JSPS Grant-in-Aid for Young Scientists (B) Grant Number 26800099.
TRS was supported by JSPS Grant-in-Aid for Young Scientists (A) Grant Number 26707007.


\appendix
\section{Virial Theorem for Clouds within Galactic Potentials}
\label{sec:VirialTheoremInGalaxies}

In order to evaluate the dynamical effects of external pressures and galactic tides 
on the dynamical states of GMCs, we present the derivation of the generalised viral equation 
for a cloud moving in a galactic potential. 
First, we introduce the moment equation of a cloud moving 
in a galactic potential in Appendix \ref{sec:App1}, 
and then we show the derivation of the viral equation of the cloud in Appendix \ref{sec:App2}.
The terms presented in Appendix \ref{sec:App2} are used for discussion of 
the effects of external pressures and galactic tides 
on the dynamical states of GMCs in Section \ref{sec:ExtendedVirialAnalysis}.

\subsection{Equations of Motion and Moment Equation}
\label{sec:App1}

In a rotating frame with an angular velocity $\Omega$, the equations of motion are given by:
\begin{eqnarray}
  \frac{{\rm D} {\bf v}_R}{{\rm D}t} = - \frac{1}{\rho}\nabla P - \nabla \Phi + {\bf a}_{\rm rot},
\end{eqnarray}
in the Lagrangian form, where $t$ is time, $D/Dt$ denotes the Lagrangian time derivative, 
${\bf v}_R$ is the velocity vector in the rotating frame, 
and $\rho$, $P$ and $\Phi$ are the mass density, thermal pressure, and gravitational potential, respectively.
The vector ${\bf a}_{\rm rot}$ is the sum of the centrifugal and Coriolis accelerations and is given by:
\begin{eqnarray}
  {\bf a}_{\rm rot} \equiv - 2{\bf \Omega} \times {\bf v}_R - {\bf \Omega} \times ({\bf \Omega} \times {\bf x}),
\end{eqnarray}
where ${\bf x}$ is the position vector.

We consider a cloud with mass $M$ and volume $V$, which rotates around the galactic centre. 
By obtaining the dot product of the equations of motion with ${\bf x}$ 
and integrating these equations over the $V$ of interest, we obtain the moment equation:
\begin{eqnarray}
 \int_V \rho {\bf x}\cdot \frac{{\rm D} {\bf v}_R}{{\rm D}t} {\rm d}V 
 = \int_V {\bf x}\cdot\left( -\frac{1}{\rho}\nabla P - \nabla \Phi + {\bf a}_{\rm rot} \right) \rho {\rm d}V.
  \label{eq:C3}
\end{eqnarray}
Following \citet{ChandrasekharFermi1953}, we can reduce the left-hand side of Eq. (\ref{eq:C3}) to:
\begin{eqnarray}
 \int_V \rho {\bf x}\cdot \frac{{\rm D} {\bf v}_R}{{\rm D}t} {\rm d}V
 = \frac{1}{2}\frac{{\rm D}^2}{{\rm D}t^2}\int_V \rho {\bf x}^2  {\rm d}V - \int_V \rho {\bf v}_R^2 {\rm d}V.
 \label{eq:C4}
\end{eqnarray}
Thus, the moment equation (Eq. (\ref{eq:C3})) can be reduced to:
\begin{eqnarray}
\label{eq:B5}
 && \frac{1}{2}\frac{{\rm D}^2}{{\rm D}t^2}\left( \int_V \rho {\bf x}^2 {\rm d}V \right) =  \nonumber\\
 && \int_V \rho {\bf v}_R^2  {\rm d}V + \int_V {\bf x}\cdot\left( -\frac{1}{\rho}\nabla P 
 	- \nabla \Phi + {\bf a}_{\rm rot} \right) \rho {\rm d}V.
\end{eqnarray}

\subsection{Virial Theorem for a Cloud Rotating in a Galaxy}
\label{sec:App2}

Introducing the centre-of-mass (CM) coordinate of a cloud, 
we can state that ${\bf x} = {\bf x}_c + {\bf r}$ and ${\bf v}_R = {\bf v}_c + {\bf u}$
 (where ${\bf x}_c$ and ${\bf v}_c$ are the position and velocity vectors of the CM, respectively, and
 ${\bf r}$ and ${\bf u}$ are the position and velocity vectors relative to the CM, respectively).
In this case, each term in equation (\ref{eq:B5}) above is reduced to:
\begin{eqnarray}
 \int_V \rho {\bf x}^2  {\rm d}V = \int_V \rho {\bf x}_c^2  {\rm d}V + \int_V \rho{\bf r}^2 {\rm d}V 
 	= M{\bf x}_c^2 + \int_V \rho{\bf r}^2  {\rm d}V, 
\end{eqnarray}
\begin{eqnarray}
 \int_V \rho {\bf v}_R^2  {\rm d}V = \int_V \rho {\bf v}_c^2  {\rm d}V + \int_V \rho {\bf u}^2 {\rm d}V
 	= M{\bf v}_c^2 + \int_V \rho {\bf u}^2  {\rm d}V, 
\end{eqnarray}
and
\begin{eqnarray}
&& \int_V {\bf x}\cdot \left( -\frac{1}{\rho}\nabla P - \nabla \Phi + 
{\bf a}_{\rm rot} \right) \rho {\rm d}V = \nonumber\\
&&	{\bf x}_c\cdot{\bf F}_c + \int_V {\bf r}\cdot \left( -\frac{1}{\rho}\nabla P - \nabla \Phi + {\bf f} \right) \rho {\rm d}V, 
\end{eqnarray}
respectively, where $M = \int_V \rho {\rm d}V$, the total force on the cloud, ${\bf F}_c$, is given by
\begin{eqnarray}
 {\bf F}_c \equiv \int_V \left[ -\frac{1}{\rho}\nabla P - \nabla \Phi 
 		-2{\bf \Omega} \times {\bf v}_c - {\bf \Omega} \times ({\bf \Omega} \times {\bf x}_c) \right]\rho {\rm d}V,
\end{eqnarray}
and ${\bf f} \equiv -2{\bf \Omega} \times {\bf u} - {\bf \Omega} \times ({\bf \Omega} \times {\bf r})$.

Substituting Eqs. (A6)--(A8) into Eq. (\ref{eq:B5}), we can obtain:
\begin{eqnarray}
&& \frac{1}{2}\frac{{\rm D}^2}{{\rm D}t^2}\left(\int_V \rho {\bf r}^2 {\rm d}V\right) 
	+ \left[\frac{1}{2}\frac{{\rm D}^2}{{\rm D}t^2}(M{\bf x}_c^2)
	- M{\bf v}_c^2 - {\bf x}_c\cdot {\bf F}_c \right]  \nonumber\\
&& = \int_V \rho {\bf u}^2  {\rm d}V + \int_V {\bf r}\cdot\left(- \frac{1}{\rho}\nabla P - \nabla \Phi + {\bf f}\right)\rho {\rm d}V.
\end{eqnarray}
Expanding the second term on the left-hand side of this equation, we can obtain:
\begin{eqnarray}
&& \frac{1}{2}\frac{{\rm D}^2}{{\rm D}t^2}(M{\bf x}_c^2) - M{\bf v}_c^2 - {\bf x}_c\cdot {\bf F}_c  \nonumber\\
&&  = (M {\bf v}_c^2 + M {\bf x}_c\cdot{\bf v}_c) - M{\bf v}_c^2 - {\bf x}_c\cdot {\bf F}_c = 0.
\end{eqnarray}
Here, we used ${\bf F}_c = M\dot{{\bf v}}_c$ in the last equation.
Thus, we can express the virial equation as:
\begin{eqnarray}
&& \frac{1}{2}\frac{{\rm D}^2}{{\rm D}t^2}\left(\int_V \rho {\bf r}^2 {\rm d}V\right) = \nonumber\\ 
&& \int_V \rho {\bf u}^2 {\rm d}V - \int_V {\bf r}\cdot\nabla P {\rm d}V - \int_V \rho{\bf r}\cdot(\nabla \Phi - {\bf f}) {\rm d}V.
\label{eq:B12}
\end{eqnarray}

The integral on the left-hand side of Eq. (\ref{eq:B12}) represents 
the moment of inertia of the cloud around the cloud's CM:
\begin{eqnarray}
 I \equiv \int_V \rho {\bf r}^2 {\rm d}V.
\end{eqnarray}
The terms on the right-hand side of Eq. (\ref{eq:B12}) have the following definitions.
\begin{itemize}
\item The first term represents twice the internal kinetic energy of the cloud, 
\begin{eqnarray}
 E_{\rm kin} \equiv \frac{1}{2} \int_V \rho {\bf u}^2 {\rm d}V.
\end{eqnarray}
\item The second term is the pressure term, which can be expressed as:
\begin{eqnarray}
 -\int_V {\bf r}\cdot \nabla P {\rm d}V 
 = -\oint_S P{\bf r}\cdot {\rm d}{\bf S} + 3 \int_V P {\rm d}V,
\end{eqnarray}
using Gauss's theorem. The first term on the right-hand side of this equation represents 
the confinement of the cloud by an external thermal pressure $P$ on the cloud boundary 
and this term is labelled as
\begin{eqnarray}
 \mathcal{T}_{\rm p} \equiv \oint_S P{\bf r}\cdot {\rm d}{\bf S}.
\end{eqnarray}
In this definition, a {\it positive} value of $\mathcal{T}_{\rm p}$ corresponds to a contribution towards the {\it collapse} of the cloud.
For a non-relativistic gas, $P$ is related to the thermal energy density, $\epsilon$, 
such that $P = 2\epsilon/3$. 
Thus, the second term can be interpreted as being twice the thermal energy of the cloud, 
$U_{\rm th} = \int_V \epsilon {\rm d}V = \frac{3}{2}\int_V P {\rm d}V$.
\item A part of the third term is the gravitational term, which can be divided into two terms:
\begin{eqnarray}
 -\int_V \rho {\bf r}\cdot \nabla \Phi {\rm d}V 
 	= - \int_V \rho {\bf r}\cdot \nabla \Phi_{\rm cl} {\rm d}V
		 - \int_V \rho {\bf r}\cdot \nabla \Phi_{\rm gal} {\rm d}V,
 \label{eq:B15}
\end{eqnarray}
where $\Phi_{\rm cl}$ and $\Phi_{\rm gal}$ are the gravitational potentials of the cloud and 
of all the mass outside the cloud (i.e. the galaxy), respectively.
The first term on the right-hand side of Eq. (\ref{eq:B15}) equates to the gravitational energy of the cloud,  
\begin{eqnarray}
 E_{\rm g} \equiv - \int_V \rho {\bf r}\cdot \nabla \Phi_{\rm cl} {\rm d}V 
 	= \frac{1}{2} \int_V \rho \Phi_{\rm cl} {\rm d}V.
\end{eqnarray}
When combined with the contribution of the centrifugal and Coriolis forces,
the second term on the right-hand side of Eq. (\ref{eq:B15}) represents the contribution of 
the tidal force to the cloud's energy budget, which is labelled as
\begin{eqnarray}
 W_{\rm tidal} \equiv - \int_V \rho {\bf r}\cdot \nabla\Phi_{\rm gal} {\rm d}V + \int_V \rho{\bf r}\cdot{\bf f} {\rm d}V,
 \label{eq:TidalTerm}
\end{eqnarray}
such that a {\it negative} value of $W_{\rm tidal}$ corresponds to a contribution towards 
the {\it collapse} of the cloud \citep{Ballesteros-Paredes2006}.
\end{itemize}
With these definitions, we can write Eq. (\ref{eq:B12}) as 
\begin{eqnarray}
 \frac{1}{2}\frac{{\rm D}^2 I}{{\rm D}t^2} 
 	= 2 (E_{\rm kin} + U_{\rm th}) - \mathcal{T}_{\rm p} + E_{\rm g} + W_{\rm tidal}. 
\end{eqnarray}

\end{document}